\documentclass[11pt,onecolumn,amssymb,nofootinbib]{revtex4} 
\usepackage{amsmath, amsthm, amscd, amssymb}
\usepackage{bm}
\usepackage{bbm}
\begin{document}

\title{\bf On the  long time behavior of stochastic
 Schr\"odinger evolutions}
\author{Angelo Bassi}
\email{bassi@ts.infn.it,  bassi@mathematik.uni-muenchen.de}
\address{Department of Theoretical Physics,
University of Trieste, Strada Costiera 11, 34014 Trieste, Italy.
\\  Mathematisches Institut der L.M.U., Theresienstr. 39, 80333
M\"unchen, Germany.}
\author{Detlef D\"urr }
\email{duerr@math.lmu.de}
\author{Martin Kolb}
\email{kolb@math.lmu.de}
\address{Mathematisches Institut der L.M.U.,
Theresienstr. 39, 80333 M\"unchen, Germany.}
\begin{abstract}
We discuss the time evolution of the wave function which is solution
of a stochastic Schr\"odinger equation describing the dynamics of a
free quantum particle subject to spontaneous localizations in space.
We prove global existence and uniqueness of solutions. Observing
that there exist three time regimes, namely the collapse regime,
after which the wave function is localized in space; the classical
regime, during which the collapsed wave function moves along a
classical path and  the diffusive regime, in which diffusion
overlaps significantly the deterministic motion we study the long
time behavior of the wave function. We assert that the general
solution converges a.s. to a diffusing Gaussian wave function having
a finite spread both in position as well as in momentum. This paper
corrects and completes earlier works on this.
\end{abstract}
\maketitle

\section{Introduction}

Stochastic differential equations (SDEs) in infinite dimensional
spaces are a subject of growing interest within the mathematical
physics and physics communities working in quantum mechanics; they
are currently used in models of spontaneous wave function
collapse~\cite{grw,grw1,grw2,pea1,pea2,pea3,di1,di2,di3,adl1,adl2,adl3,bas1,bas2},
in the theory of continuous quantum
measurement~\cite{bel1,bel2,bel3,bar1,bar2,bar3,bla1,bla2,bla3}, and
in the theory of open quantum systems~\cite{dec1,dec2,dec3}. In the
first case, the Schr\"odinger equation is modified by adding
appropriate non-linear and stochastic terms which induce the
(random) collapse of the wave function in space; in this way, one
achieves the goal of a unified description of microscopic quantum
phenomena and macroscopic classical ones, avoiding the occurrence of
macroscopic quantum superpositions. In the second case, using  the
projection postulate,  stochastic terms in the Schr\"odinger
equation are used to describe the effect of a continuous
measurement. In the third case, slightly generalising the notion of
continuous measurement to generic interactions with environments,
SDEs are used as phenomenological equations describing the
interaction of a quantum system with an environment, the stochastic
terms encoding the effect of the environment on the system. Looking
directly at the stochastic differential equation for the wave
function, rather than  the deterministic equation of the Lindblad
type for the statistical operator has some advantages with respect
to the standard master equation approach, e.g. for faster numerical
simulations~\cite{pet}.

Among the different SDEs which have been considered so far, the
following equation, defined in the Hilbert space ${\mathcal H}
\equiv {\mathcal L}^2({\mathbb R})$, is of particular
interest~\cite{gg,hol,bel2,kol1,kol2,kol3,kol4,kol5,alb,ab1,ch}:
\begin{equation} \label{eq:non-lin}
d \psi_{t} \; = \; \left[ - \frac{i}{\hbar}\, \frac{p^2}{2m} \, dt
\; + \; \sqrt{\lambda}\, ( q - \langle q \rangle_{t} )\, dW_{t} \; -
\; \frac{\lambda}{2}\, ( q - \langle q \rangle_{t} )^2\, dt
\right]\, \psi_{t}, \qquad\qquad \psi_{0} = \psi.
\end{equation}
The first term on the right-hand-side represents the usual quantum
Hamiltonian of a free particle in one dimension, $p$ being the
momentum operator. The second and third terms of the equation, as we
shall see, induce the localization of the wave function in space;
$q$ is the position operator and $\langle q \rangle_{t}$ denotes the
quantum expectation $\langle \psi_{t} | q \, \psi_{t}\rangle$ of $q$
with respect to $\psi_{t}$. The parameter $\lambda$ is a fixed
positive constant which sets the strength of the collapse mechanism,
while $W_{t}$ is a standard Wiener process defined on a probability
space $(\Omega, {\mathcal F}, {\mathbb P})$ with filtration $\{
{\mathcal F}_{t}, t \geq 0 \}$.

Eq.~\eqref{eq:non-lin} plays a special role among the SDEs in
Hilbert spaces because it is the simplest exactly solvable equation
describing the time evolution of a non-trivial  physical system.
Within the theory of continuous quantum measurement, it describes a
measurement-like process designed to measure the position of a free
quantum particle; within decoherence theory  it represents one of
the possible unravellings of the master equation first derived by
Joos and Zeh~\cite{jz}. Within collapse models (like GRW-models), it
may describe the evolution of a free quantum particle (or the center
of mass of an isolated system) subject to spontaneous localizations
in space ~\cite{grw},~\cite{grw1} in the following sense. Realistic
models of spontaneous wave function collapse are based on a more
complicated stochastic differential equation: The difference between
Eq.~\eqref{eq:non-lin} and the equations of the standard
localization models such as GRW~\cite{grw} and CSL~\cite{grw1} is
most easily described on the level of the Lindblad equations for the
respective  statistical operators $\rho_t := {\mathbb E}_{\mathbb P}
[|\psi_t \rangle \langle \psi_t |]$,  induced by the stochastic
dynamics of the wave function. By virtue of  Eq.~\eqref{eq:non-lin}
(see e.g. ~\cite{di3}):
\begin{equation} \label{eq:dfxds}
\frac{d}{dt}\, \rho_t \;  = \; -\frac{i}{2m\hbar}\, [ p^2, \rho_t ]
\; - \; \frac{\lambda}{2}\, [q, [q, \rho_t ]],
\end{equation}
with  the ``Lindblad term''  in position representation
\begin{equation}\frac{\lambda_{\makebox{\tiny GRW}}\alpha}{4}
(x-y)^2\rho_t(x,y)\,.
\end{equation}
 For
the GRW dynamics as described in~\cite{grw} the corresponding
Lindblad term of the GRW master equation in the position
representation reads:
\begin{equation}
- \lambda_{\makebox{\tiny GRW}} \left[ 1 - e^{-\alpha (x - y)^2/4}
\right]\rho_t(x,y)\,.
\end{equation}
When the distances involved are smaller than the length
$1/\sqrt{\alpha}\simeq 10^{-5}$ cm characterizing the model we have
that
\begin{equation}
- \lambda_{\makebox{\tiny GRW}} \left[ 1 - e^{-\alpha (x - y)^2/4}
\right] \; \simeq \; \frac{\lambda_{\makebox{\tiny GRW}}\alpha}{4}
(x-y)^2 \qquad \makebox{for: $ |x-y| \ll 1/\sqrt{\alpha}$}\,.
\end{equation}
 Accordingly, the stochastic dynamics of
Eq.~\eqref{eq:non-lin}  approximates--at least on the statistical
level--the GRW dynamics  for all atomic and subatomic distances.
Since this is a regime of growing interest ~\cite{bo1,bo2,bo3,ad} it
is reasonable to study now first the simpler equation
Eq.~\eqref{eq:non-lin}.

Eq.~\eqref{eq:non-lin} is non-linear. Non-linearity is a fundamental
ingredient because only in this way it is possible to reproduce the
collapse of the wave function. It is well known how to ``linearize''
the equation, i.e. how to express its solutions as a function of the
solutions of a suitable linear SDE~\cite{hol,bar}. We briefly review
this procedure.

Let us consider the following linear SDE:
\begin{equation} \label{eq:lin}
d \phi_{t} \; = \; \left[ - \frac{i}{\hbar}\, \frac{p^2}{2m} \, dt
\; + \; \sqrt{\lambda}\, q\, d\xi_{t} \; - \; \frac{\lambda}{2}\,
q^2\, dt \right] \phi_{t}, \qquad\qquad \phi_{0} = \phi,
\end{equation}
defined in the same Hilbert space ${\mathcal H} \equiv {\mathcal
L}^2({\mathbb R})$; the stochastic process $\xi_{t}$ is a standard
Wiener process with respect to the probability space $(\Omega,
{\mathcal F}, {\mathbb Q})$ and filtration $\{ {\mathcal F}_{t}, t
\geq 0 \}$, where $\mathbb{Q}$ is a new probability measure whose
relation with ${\mathbb P}$ will soon be established. This equation
does not conserve the norm of the state vector, as the evolution is
not unitary; we therefore introduce the normalized state vectors:
\begin{equation} \label{eq:step1}
\psi_{t} \; = \; \left\{
\begin{array}{ll}
\phi_{t}/\|\phi_{t}\| \quad\quad & \makebox{if: $\|\phi_{t}\| \neq 0$}, \\
0 & \makebox{otherwise};
\end{array}
\right.
\end{equation}
A standard application of It\^o calculus shows that, if $\phi_{t}$
solves Eq.~\eqref{eq:lin}, then $\psi_{t}$ defined
in~\eqref{eq:step1} solves the following non-linear SDE:
\begin{equation} \label{eq:non-lin-1}
d \psi_{t} \; = \; \left[ - \frac{i}{\hbar} \frac{p^2}{2m} dt \, +
\, \sqrt{\lambda} ( q - \langle q \rangle_{t} ) (d\xi_{t} - 2
\sqrt{\lambda} \langle q \rangle_{t}dt) \, - \, \frac{\lambda}{2} (
q - \langle q \rangle_{t} )^2 dt \right] \psi_{t},
\end{equation}
for the same initial condition $\psi = \phi$.

Eq.~\eqref{eq:non-lin-1} is a well defined collapse equation,
however it is not suitable for physical applications, as the
collapse does not occur with the correct quantum probabilities. This
can be seen by analyzing the time evolution of particular solutions,
such as Gaussian wave functions; it can also be easily understood by
noting that there is no fundamental difference between
Eq.~\eqref{eq:non-lin-1} and Eq.~\eqref{eq:lin}, since any solution
of Eq.~\eqref{eq:non-lin-1} can be obtained from a solution of
Eq.~\eqref{eq:lin} simply by normalizing the wave function. In turn,
Eq.~\eqref{eq:lin} does not contain any information as to why the
wave function should collapse according to the Born probability
rule, i.e. the Wiener process $\xi_{t}$ is not forced to pick most
likely those values necessary to reproduce quantum probabilities,
during the collapse process.

The way to include such a feature into the dynamical evolution of
the wave function is to replace the measure $\mathbb{Q}$ with a new
measure (which will turn out to be the measure $\mathbb{P}$
previously introduced) so that the process $\xi_{t}$, according to
the new measure, is forced to take with higher probability the
values which account for quantum probabilities. This is precisely
the key idea behind the original GRW model of spontaneous wave
function collapse~\cite{grw}: the wave function is more likely to
collapse where it is more appreciably different from zero. The
mathematical structure of the GRW model suggests that the square
modulus $\| \phi_{t} \|^2$ should be used as density for the change
of measure. We now formalize these steps.

In~\cite{hol}, Holevo has proven that $\| \phi_{t} \|^2$ is a
martingale satisfying the equation:
\begin{equation} \label{eq:hfgdsd4}
\|\phi_{t}\|^2 \; = \; \|\phi_{0}\|^2 + 2 \sqrt{\lambda}
\int_{0}^{t} \langle q \rangle_{s} \|\phi_{s}\|^2 d \xi_{s};
\end{equation}
when $\| \phi_{0} \|^2 = 1$, and from now on we will always assume
that this is the case, $\| \phi_{t} \|^2$ can be used as a
Radon-Nikodym derivative to generate a new probability measure
${\mathbb P}$ from ${\mathbb Q}$, according to the usual formula:
\begin{equation} \label{eq:xgvcvc}
{\mathbb P}[E] \; := \; {\mathbb E}_{\mathbb Q}[1_{E}\|\phi_{t}\|^2]
\qquad \forall \,\, E \, \in \, {\mathcal F}_{t} \quad \forall \,\,
t < +\infty,
\end{equation}
where $1_{E}$ is the indicator function relative to the measurable
subset $E$. The martingale property, together with the property
${\mathbb E}_{\mathbb Q}[\|\phi_{t}\|^2] = 1$, guarantee consistency
among different times, so that~\eqref{eq:xgvcvc} defines a unique
probability measure ${\mathbb P}$. In the following, for simplicity
we will write $d {\mathbb P} / d {\mathbb Q} \equiv \|\phi_{t}\|^2$.
One can then show that Eq.~\eqref{eq:non-lin-1}, with the stochastic
dynamics defined on the probability space $(\Omega, {\mathcal F},
{\mathbb P})$ in place of $(\Omega, {\mathcal F}, {\mathbb Q})$,
correctly describes the desired physical situations.

A drawback of the change of measure is that the equation is defined
in terms of the stochastic process $\xi_{t}$, which is not anymore a
Wiener proves with respect to the measure ${\mathbb P}$, as it was
with respect to the measure ${\mathbb Q}$. This can be a source of
many difficulties, e.g. when analyzing the properties of the
solutions of the equation. The disadvantage can be removed by
resorting to Girsanov's theorem, which connects Wiener processes
defined on the same measurable space, but with respect to different
probability measures. According to this theorem, the process
\begin{equation} \label{eq:step2}
W_{t} \quad := \quad \xi_{t} \; - \; 2\sqrt{\lambda} \int_{0}^{t}
\langle q \rangle_{s} \, ds,
\end{equation}
is a Wiener process with respect to $(\Omega, {\mathcal F}, {\mathbb
P})$ and filtration $\{ {\mathcal F}_{t}, t \geq 0 \}$, and thus is
the natural process for describing the stochastic dynamics with
respect to the measure ${\mathbb P}$. It is immediate to see that,
once written in terms of $W_{t}$, Eq.~\eqref{eq:non-lin-1} reduces
to Eq.~\eqref{eq:non-lin}, thus the link between Eq.~\eqref{eq:lin}
and~\eqref{eq:non-lin} is established. The above discussion should
also have given a first idea of why SDEs like Eq.~\eqref{eq:non-lin}
are those which are used in Quantum Mechanics to described the
collapse of the wave function; we will come back on this point later
in the paper.

The first important problem to address concerns the status of the
solutions of Eq.~\eqref{eq:lin}. In~\cite{hol}, Holevo has proven
the existence and uniqueness of  topological weak solutions of a
rather general class of SDEs with unbounded operators, to which
Eq.~\eqref{eq:lin} belongs. (See the end of the section for the
notation.) The problem of the existence and uniqueness of
topological strong solutions of Eq.~\eqref{eq:lin} has been
addressed in~\cite{gg}; there however, the proof relies on the
expansion of wave functions in terms of Gaussian states, which in
general is problematic and requires special care, as shown
in~\cite{epl}. An explicit representation of the strong strong
solution of Eq.~\eqref{eq:lin} has been given in~\cite{alb}; the
representation is written in terms of path integrals and is not
particularly suitable for analyzing the time evolution of the
general solution. A much more convenient representation, given in
terms of the Green's function of Eq.~\eqref{eq:lin}, has been first
derived in~\cite{kol1,kol4}; the Green's function reads:
\begin{equation} \label{eq:dfvc}
G_{t}(x,y) \; = \; K_{t} \exp \left[ - \frac{\alpha_{t}}{2}(x^2 +
y^2) + \beta_{t} x y  \, + \, \overline{a}_{t}x \, + \,
\overline{b}_{t}y \, + \, \overline{c}_{t}\right];
\end{equation}
the coefficients $K_{t}$, $\alpha_{t}$ and $\beta_{t}$ are
deterministic and equal to
\begin{eqnarray}
K_{t} & = &  \sqrt{\frac{\lambda}{\upsilon \pi \sinh \upsilon t}},
\label{eq:sad1} \\
\alpha_{t} & = & \frac{2\lambda}{\upsilon} \coth \upsilon t, \label{eq:sad2}\\
\beta_{t} & = & 2\, \frac{\lambda}{\upsilon}\, \sinh^{-1} \!\upsilon
t, \label{eq:sad3}
\end{eqnarray}
while the remaining coefficients are functions of the Wiener process
$\xi_{t}$:
\begin{eqnarray}
\overline{a}_{t} & = & \sqrt{\lambda} \sinh^{-1} \!\upsilon t
\int_{0}^{t} \sinh \upsilon s\, d \xi_{s}, \label{eq:sad4}\\
\overline{b}_{t} & = & \frac{2i\hbar}{m} \frac{\lambda}{\upsilon}
\int_{0}^{t} \frac{\overline{a}_{s}}{\sinh \upsilon s}\, ds, \label{eq:sad5}\\
\overline{c}_{t} & = & \frac{i\hbar}{m} \int_{0}^{t}
\overline{a}_{s}^{2} \, ds. \label{eq:sad6}
\end{eqnarray}
In the above expressions, we have introduced the following two
constants:
\begin{equation} \label{eq:gfdsp}
\upsilon \; \equiv \; \frac{1+i}{2}\, \omega, \qquad\quad \omega \;
\equiv \; 2\, \sqrt{\frac{\hbar \lambda}{m}}.
\end{equation}
As we shall see, the parameter $\omega$, which has the dimensions of
a frequency, will set the time scales for the collapse of the wave
function. The representation in terms of the Green's
function~\eqref{eq:dfvc}, as we said, is particularly suitable for
analyzing the time evolution of the general solution of
Eq.~\eqref{eq:lin}, and thus of Eq.~\eqref{eq:non-lin}, even though
we will see that, when studying the long time behavior, another
representation is more convenient.

Our first result concerns the meaning  of the solution of
Eq.~\eqref{eq:lin} in terms of  \begin{equation} \label{eq:fbxc}
\phi_{t}(x) \; := \; \int dy\, G_{t}(x,y)\, \phi(y)
\end{equation}for  given initial
condition $\phi$.\vskip 0.3cm

\noindent \textsc{Theorem 1 (Solution):} let $\phi_{t}$ be defined
as in~\eqref{eq:fbxc}; then the following three statements hold true
with ${\mathbb Q}$-probability 1:
\begin{eqnarray}
1. & & \phi \, \in \, {\mathcal L}^{2}({\mathbb R}) \,\quad
\Rightarrow \quad \phi_{t} \, \in \, {\mathcal L}^{2}({\mathbb R}),
\\
2. & & \phi \, \in \, {\mathcal L}^{2}_{B}({\mathbb R}) \quad
\Rightarrow \quad \phi_{t} \; \text{is a topological strong solution of Eq.~\eqref{eq:lin}},\\
3. & & \phi \, \in \, {\mathcal L}^{2}({\mathbb R}) \quad
\Rightarrow \quad \lim_{t \rightarrow 0} \| \phi_{t} - \phi \| \, =
0,
\end{eqnarray}
where ${\mathcal L}^{2}_{B}({\mathbb R})$ is the subspace of all
{\it bounded} functions of ${\mathcal L}^{2}({\mathbb R})$. \vskip
0.3cm

Having the explicit solution of the Eq.~\eqref{eq:lin}, and thus of
Eq.~\eqref{eq:non-lin}, the next relevant problem is to unfold its
physical content. Previous analysis of similar
equations~\cite{grw1,di2,adl1,bas2,ab1} have shown that one can
identify three regimes, which are more or less well separated
depending on the value of the parameters $\lambda$ and $m$.
\begin{enumerate}
\item
  {\it Collapse regime}: A wave function having an initial
large spread, localizes in space, the localization occurring in
agreement with the Born probability rule.\item {\it Classical
regime:} The localized wave function moves in space like a classical
free particle, since the fluctuations due to the Wiener process can
be safely ignored. \item {\it
 Diffusive regime:} Eventually, the
random fluctuations become dominant and the wave function starts to
diffuse appreciably. \end{enumerate}

It is not an easy task to spell out rigorously these regimes and
their properties. We shall however be a bit more specific on this in
the following section. We shall afterwards focus on the simplest
regime, namely the diffusive one, which in fact has been intensively
looked at in the previous years~\cite{di1,bel3,dec1,kol3,kol4,ab1}
and we shall prove a remarkable property of the solutions of
Eq.~\eqref{eq:non-lin}: Any solution converges almost surely to a
Gaussian state wave function having a fixed spread. \vskip 0.3cm

\noindent \textsc{Theorem 2 (Large time behavior):} let $\psi_{t}$
be a solution of Eq.~\eqref{eq:non-lin}; then under conditions which
we will specify, the following property holds true with ${\mathbb
P}$-probability 1:
\begin{equation}
\lim_{t \rightarrow \infty} \| \psi_{t} - \psi^{\text{\tiny
$\infty$}}_{t} \| \; = \; 0,
\end{equation}
where $\psi^{\text{\tiny $\infty$}}_{t}$, defined
in~\eqref{eq:dgh5}, is a Gaussian wave function with a fixed spread
both in position and momentum. \vskip 0.3cm

Theorem 1 and Theorem 2 have been extensively discussed before in
the literature ~\cite{di1,bel3,dec1,gg,kol3,kol4,kol5,ab1}, proving
that the community has devoted much attention to the problem.
However, these proofs are not complete or flawed. Concerning Theorem
1, in particular Statement 3 was not
proven~\cite{gg,kol3,kol4,kol5}. While Statements 1 and 2 are rather
straightforward conclusions from the Gaussian kernel of the
propagator, the third Statement is much more subtle and does not
follow from purely analytical arguments. Concerning Theorem 2, none
of the previous proofs is decisive. In~\cite{kol4,kol5}, the major
flaw was that it was overlooked that the eigenfunction expansion of
the relevant dissipative operator (not self-adjoint) does not give
rise to an orthonormal basis. In~\cite{bel3}, the long time behavior
was analyzed by expanding the general solution in terms of coherent
states, while in~\cite{dec1,ab1} it was analyzed by scrutinizing the
time evolution of the spread in position of the solution;
in~\cite{epl} it has been shown that both approaches are not
conclusive. Finally,~\cite{di1} proposed Theorem 2 as a conjecture,
but shows stability of $\psi^{\text{\tiny $\infty$}}_{t}$ only
against small perturbations. Building on previous work of Holevo,
Mora and Rebolledo recently enhanced in \cite{Reb1} and \cite{Reb2}
the general theory of stochastic Schr\"odinger equations. In
particular they developed criteria for the existence of regular
invariant measures for a large class of stochastic Schr\"odinger
equations as an important step towards an understanding of the large
time behavior. Until now however the only complete and detailed
results on  the large time behavior seem to be  Theorems 1 and
2.

We conclude this introductory section by summarizing the content of
the paper. In Sec.~\ref{sec:TimeEvo} we will present a qualitative
analysis of the time evolution of the general solution of
Eq.~\eqref{eq:non-lin}; we will discuss the three regimes previously
introduced, giving also numerical estimates, and we will set the
main problems which we aim at solving. In Sec.~\ref{sec:two} we will
analyze the structure of the Green's function~\eqref{eq:dfvc} and
prove theorem 1. In Sec.~\ref{sec:three} we will introduce another
representation of the general solution of Eq.~\eqref{eq:non-lin},
which is more suitable for analyzing its long time behavior.
Sec.~\ref{sec:four} will be devoted to the proof of theorem 2.
Finally, Sec.~\ref{sec:end} will contain some concluding remarks and
an outlook. \vskip 0.3cm

\noindent \textsc{Notation.} We will work in the complex and
separable Hilbert space ${\mathcal L}^{2}({\mathbb R})$, with the
norm and the scalar product given, respectively, by $\| \cdot \|$
and $\langle \cdot | \cdot \rangle$. We will also consider the
subspace ${\mathcal L}^{2}_{B}({\mathbb R})$ of all {\it bounded}
functions of ${\mathcal L}^{2}({\mathbb R})$. Given an operator $O$,
we denote with ${\mathcal D}(O)$ its domain and with ${\mathcal
R}(O)$ its range.

Since in some expression the real and imaginary parts of some
coefficients appear, we introduce for ease of readability the
symbols $z^{\makebox{\tiny R}}$ or $z_{\makebox{\tiny R}}$ will
denote the real part of the complex number $z$, while
$z^{\makebox{\tiny I}}$ or $z_{\makebox{\tiny I}}$ will denote its
imaginary part.

Given the linear SDE~\eqref{eq:lin}, a {\it topological strong
solution} is an ${\mathcal L}^2$-values process such that for any
$t>0$,
\begin{equation}
\phi_t \; = \; \phi \, - \, \frac{i}{\hbar} \int_0^t
\frac{p^2}{2m}\phi_s ds \, + \, \sqrt{\lambda} \int_0^t q \phi_s
dW_s \, - \, \frac{\lambda}{2} \int_0^t q^2 \phi_s ds
\end{equation}
holds with ${\mathbb Q}$-probability 1. A {\it topological weak
solution} instead is an ${\mathcal L}^2$-values process such that
for any $t>0$ and for any $\chi \in {\mathcal D}(p^2)\cap {\mathcal
D}(q^2)$,
\begin{equation}
\langle \chi | \phi_t \rangle \; = \; \langle \chi | \phi \rangle \,
- \, \frac{i}{\hbar} \int_0^t \frac{1}{2m} \langle p^2 \chi |\phi_s
\rangle ds \, + \, \sqrt{\lambda} \int_0^t \langle q \chi | \phi_s
\rangle dW_s \, - \, \frac{\lambda}{2} \int_0^t \langle q^2 \chi |
\phi_s \rangle ds
\end{equation}
holds with ${\mathbb Q}$-probability 1. Topological strong and week
solutions for the nonlinear SDE~\eqref{eq:non-lin} are defined in a
similar way.

There is also a distinction between {\it strong} and {\it weak}
solutions in a {\it stochastic} sense~\cite{ls}, depending on
whether the probability space, the filtration and the Wiener process
are given {\it a priori} (strong solution) or whether they can be
constructed in such a way to solve the required SDE (weak solution).
Throughout the paper we will deal only with strong solutions in the
stochastic sense.


\section{Time evolution of the general solution}
\label{sec:TimeEvo}

We begin our discussion with a qualitative analysis of the time
evolution of the general solution of Eq.~\eqref{eq:non-lin}; we will
spot out the regimes we introduced in the previous section,
corresponding to three different behaviors of the wave function.
These regimes of course depend on the value of the mass $m$ of the
particle and also on the value of the coupling constant $\lambda$
which sets the strength of the collapse mechanism. As discussed e.g.
in~\cite{ab1}, it is physically appropriate to take $\lambda$
proportional to the mass $m$ according to the formula:
\begin{equation}
\lambda \; := \; \lambda_{0} \, \frac{m}{m_{0}},
\end{equation}
where $\lambda_{0}$ is now assumed to be a {\it universal} coupling
constant, while $m_{0}$ is taken equal to the mass of a nucleon
($\simeq 1.67 \times 10^{-27}$ kg). To be definite, in the following
we take $\lambda_{0} \simeq 1.00 \times 10^{-2}$ m$^{-2}$
sec$^{-1}$, so that the localization mechanism has the same strength
as that of the GRW model~\cite{grw}. Though, as we discussed in the
introduction, Eq.~\eqref{eq:non-lin} is used also in the context of
the theory of continuous measurement as well as in the theory of
decoherence, for brevity and clarity in the following we will only
make reference to its application within models of spontaneous wave
function collapse. \vskip 0.3cm

\noindent \textsc{1. The collapse regime.} The first important
effect of the dynamics embodied in Eq.~\eqref{eq:non-lin} is that a
wave function, which initially is well spread out in space, becomes
rapidly localized. This is most easily seen through the Green's
function representation of the solution. The Green's function
$G_{t}(x,y)$ in~\eqref{eq:dfvc} can be rewritten as follows
\begin{equation}
G_{t}(x,y) \; = \; K_{t}\, \exp \left[ -
\frac{\tilde{\alpha}_{t}}{2}\, x^2 + \tilde{a}_{t} x + \tilde{c}_{t}
\right] \, \exp \left[ - \frac{\alpha_{t}}{2}\, (y - Y^{x}_{t})^2
\right]
\end{equation}
where we have introduced the new parameters:
\begin{eqnarray}
\tilde{\alpha}_{t} & = & \alpha_{t} -
\frac{\beta^{2}_{t}}{\alpha_{t}} \; = \; \frac{2
\lambda}{\upsilon}\, \tanh \upsilon t, \\
\tilde{a}_{t} & = & \overline{a}_{t} +
\frac{\beta_{t}\overline{b}_{t}}{\alpha_{t}}, \\
\tilde{c}_{t} & = & \overline{c}_{t} +
\frac{\overline{b}_{t}^2}{2\alpha_{t}}, \\
Y^{x}_{t} & = & \frac{\beta_{t} x + \overline{b}_{t}}{\alpha_{t}}.
\end{eqnarray}
The $y$-part of $G_{t}(x,y)$ is a Gaussian function whose spread in
position (equal to $1/\sqrt{\alpha^{\text{\tiny R}}_{t}})$ rapidly
decreases in time, and afterwards remains very small. In particular,
we have:
\begin{equation} \label{eq:hdgsf}
\alpha^{\text{\tiny R}}_{t} \; = \; \frac{2\lambda}{\omega}
\frac{\sinh \omega t - \sin \omega t}{\cosh \omega t - \cos \omega
t} \; = \; \left\{
\begin{array}{llll}
\displaystyle \frac{2}{3}\, \lambda t & \simeq & (3.99 \times
10^{24} \text{m$^{-2}$ Kg$^{-1}$ sec$^{-1}$}) m t &\qquad t \ll
\omega^{-1}, \\
& & & \\ \displaystyle \frac{2\lambda}{\omega } & \simeq & (2.39
\times 10^{29} \text{m$^{-2}$ Kg$^{-1}$}) m &\qquad t \rightarrow +
\infty,
\end{array}
\right.
\end{equation}
with $\omega \simeq 5.01 \times 10^{-5}$ sec$^{-1}$ independent of
the mass of the particle.

Let us introduce a length $\ell$, and let say that a wave function
is {\it localized} when its spread is smaller than $\ell$. For sake
of definiteness, we take $\ell \simeq 1.00 \times 10^{-7}$ m,
corresponding to the width of the collapsing Gaussian of the GRW
model. By means of this length, we can define the {\it collapse
time} $t_{1}$ as the time when the spread of the $y$-part of the
Green's function $G_{t}(x,y)$ becomes smaller than $\ell$. By using
the small time approximation of $\alpha^{\text{\tiny R}}_{t}$ given
in~\eqref{eq:hdgsf}, we can set:
\begin{equation}
t_{1} \; := \; \frac{3}{2\ell^2 \lambda} \; \simeq \; \frac{2.51
\times 10^{-11}\, \text{Kg sec}}{m}.
\end{equation}
As we see, and as we expect, this time decreases for increasing
masses, i.e. for increasing values of $\lambda$, and is very small
for macroscopic particles.

Let us assume that the initial state $\phi(x)$ is not already
localized, and in particular that it does not change appreciably on
the scale set by $\ell$; this is a physically reasonable assumption
when $\phi$ represents the state of the center of mass of a
macroscopic object. In this case, from the time $t_{1}$ on, the
$y$-part of the Green's function $G_{t}(x,y)$ acts like a
Dirac-delta on $\phi(x)$, and the solution at time $t$ of the linear
equation can be written as follows:
\begin{equation}
\phi_{t}(x) \; \simeq \; \sqrt{\frac{2\pi}{\alpha_{t}}}\, K_{t} \exp
\left[ - \frac{\tilde{\alpha}_{t}}{2}\, x^2 + \tilde{a}_{t} x +
\tilde{c}_{t} \right] \, \phi(Y^{x}_{t});
\end{equation}
This is a Gaussian state whose spread is controlled by
$\tilde{\alpha}_{t}$, which evolves in time in a way similar to
$\alpha_{t}$; in particular:
\begin{equation} \label{eq:hdhcgsf}
\tilde{\alpha}^{\text{\tiny R}}_{t} \; = \; \frac{2\lambda}{\omega}
\frac{\sinh \omega t + \sin \omega t}{\cosh \omega t + \cos \omega
t} \; = \; \left\{
\begin{array}{llll}
\displaystyle 2 \lambda t & \simeq & (1.20 \times 10^{25}
\text{m$^{-2}$ Kg$^{-1}$ sec$^{-1}$}) m t &\qquad t \ll
\omega^{-1}, \\
& & & \\ \displaystyle \frac{2\lambda}{\omega } & \simeq & (2.39
\times 10^{29} \text{m$^{-2}$ Kg$^{-1}$}) m &\qquad t \rightarrow +
\infty.
\end{array}
\right.
\end{equation}
As we see, the spread $1/\sqrt{\tilde{\alpha}^{\text{\tiny R}}_{t}}$
is well below $\ell$, for any $t \geq t_{1}$. We can the conclude
that, for times greater than the collapse time, any state initially
well spread out in space is mapped into a very well localized wave
function.

An important issue is {\it where} the wave function collapses to,
given that the initial state is spread out in space. We now show
that the position of the wave function after the collapse is
distributed in very good agreement with the Born probability rule.

A reasonable measure of where the wave function is, after it has
collapsed, is given by the quantum average of the position operator
$\langle q \rangle_{t}$. Accordingly, the probability for the
collapsed wave function to lie within a Borel measurable set $A$ of
${\mathbb R}$ can be simply defined to be ${\mathbb P}^{\text{\tiny
coll}}_{t}[A] := {\mathbb P}[\omega : \langle q \rangle_{t} \in A]$.
Though this probability is mathematically well defined for any Borel
measurable subset $A$, it is physically meaningful only when $A$
represents an interval $\Delta$ much larger than the spread of the
wave function itself, or a sum of such intervals. In such a case, as
discussed in~\cite{bgs}, one can show that:
\begin{equation} \label{qe:iox}
{\mathbb P}^{\text{\tiny coll}}_{t}[A] \; \simeq \; {\mathbb
E}_{\mathbb P} [ \| P_{\Delta} \psi_{t} \|^2 ] \; \equiv \;
\int_{\Delta} p_{t}(x) dx,
\end{equation}
where $P_{\Delta}(x)$ is the characteristic function of the interval
$\Delta$ of the real axis and $p_{t} = {\mathbb E}_{\mathbb P} [
|\psi_{t}(x)|^2]$. The idea behind the approximate
equality~\eqref{qe:iox} is that when $\psi_{t}$ lies within
$\Delta$, then $P_{\Delta} \psi_{t} \simeq \psi_{t}$, so that $\|
P_{\Delta} \psi_{t} \|^2$ is almost equal to 1, while when it lies
outside $\Delta$, it is practically 0. The critical situations,
which require special care, are those when the wave function lies at
the edges of $\Delta$.

In~\cite{ab1} it has been proven that:
\begin{equation} \label{eq:vbvb}
p_{t}(x) \; = \; \sqrt{\frac{\mu_{t}}{\pi}} \int dy\, e^{- \mu_{t}
y^2} p^{\text{\tiny Sch}}_{t} (x+y), \qquad \mu_{t} = \frac{3 m
m_{0}}{2 \hbar^2 \lambda_{0}t^3} \simeq (2.27 \times 10^{43}
\text{m$^{-2}$ Kg$^{-1}$ sec$^{3}$})\, \frac{m}{t^3},
\end{equation}
where $p^{\text{\tiny Sch}}_{t}(x) = |\psi^{\text{\tiny
Sch}}_{t}(x)|^2$ and $\psi^{\text{\tiny Sch}}_{t}(x)$ is the
solution of the standard free-particle Sch\"odinger equation, for
the given initial condition $\phi(x)$. For the times we are
considering ($t = t_{1}$), the Gaussian term in~\eqref{eq:vbvb} is
much more peaked than any typical quantum probability distribution
$p^{\text{\tiny Sch}}_{t}(x)$, and consequently acts like a
Dirac-delta on it; accordingly, $p_{t}(x) \simeq p^{\text{\tiny
Sch}}_{t}(x)$. Finally, for macroscopic systems and for the times we
are considering, the wave function solution of the free-particle
Schr\"odinger equation does not change appreciably, implying that
$p^{\text{\tiny Sch}}_{t}(x) \simeq p^{\text{\tiny Sch}}_{0}(x) =
|\phi(x)|^2$, which means precisely that the collapse probability is
distributed in agreement with the Born probability rule. \vskip
0.3cm

\noindent \textsc{2. The classical regime.} After time $t_{1}$, we
are left with a wave function which, when $m$ is the mass of a
macroscopic particle, is very well localized in space, almost
point-like. This is the way in which collapse model reproduce the
particle-like behavior of classical systems, within the framework of
a wave-like dynamics. The relevant question now is to unfold the
time evolution of the position and momentum of the wave function, to
see whether it matches Newton's laws.

When the wave function is well localized in space ($t > t_{1}$), one
can reasonably assume that it can be approximated with the Gaussian
state to which---as we shall see---it asymptotically converges to.
We will analyze the time evolution of such a Gaussian state in the
following, and we will see that its mean position $\overline{x}_{t}$
and momentum $\hbar \overline{k}_{t}$ evolve in time as follows (see
Eqs.~\eqref{eq:prima1} and~\eqref{eq:seconda1}):
\begin{eqnarray}
\overline{x}_{t} & = & \overline{x}_{t_{1}} + \frac{\hbar}{m}
\overline{k}_{t_{1}} (t - t_{1}) + \sqrt{\lambda} \frac{\hbar}{m}
\int_{t_{1}}^{t} W_{s} ds + \sqrt{\frac{\hbar}{m}}
(W_{t} - W_{t_{1}}),\label{eq:bvxcdp}\\
\overline{k}_{t} & = & \overline{k}_{t_{1}} + \sqrt{\lambda} (W_{t}
- W_{t_{1}}).
\end{eqnarray}
We can easily recognize in the deterministic parts of the above
equations the free-particle equations of motions of classical
mechanics describing a particle moving along a straight line with
constant velocity; the remaining terms are the fluctuations around
the classical motion, driven by the Brownian motion $W_{t}$. The
important feature of the above equations is that these fluctuations,
for macroscopic masses, are very small, for very long times. As a
matter of fact, if we estimate the Brownian motion fluctuations by
setting $W_{t} \sim \sqrt{t}$, we have for the stochastic terms in
Eq.~\eqref{eq:bvxcdp}:
\begin{eqnarray}
\sqrt{\lambda} \frac{\hbar}{m} \int_{t_{1}}^{t} W_{s} ds & \simeq &
\frac{2}{3} \sqrt{\lambda} \frac{\hbar}{m}\, t^{3/2} \; \simeq \;
(1.63 \times 10^{-22}
\text{m Kg$^{1/2}$ sec$^{-3/2}$})\, \frac{t^{3/2}}{\sqrt{m}}, \label{eq:dsfv}\\
\sqrt{\frac{\hbar}{m}} (W_{t} - W_{t_{1}}) & \simeq &
\sqrt{\frac{\hbar\, t}{m}} \;\;\;\;\;\;\quad\; \simeq \; (1.02
\times 10^{-17} \text{m Kg$^{1/2}$ sec$^{-1/2}$})\,
\sqrt{\frac{t}{m}}. \label{eq:dsfv2}
\end{eqnarray}
We see that the random fluctuations decrease with the square root of
the mass $m$ of the particle, which means that the bigger the
system, the more deterministic its motion. This is how collapse
models recover classical determinism at the macroscopic level, from
a fundamentally stochastic theory.

We can introduce a time $t_{2}$, defined as the time after which the
fluctuations become larger than $L$; we can set e.g. $L \simeq 1.00
\times 10^{-3}$ m. Since the fluctuations in~\eqref{eq:dsfv} grow
faster as those in~\eqref{eq:dsfv2}, we can set:
\begin{equation}
t_{2} \; \simeq \; \left(
\frac{3}{2}\frac{L}{\sqrt{\lambda}}\frac{m}{\hbar} \right)^{2/3} \;
\simeq \; (3.55 \times 10^{12}\, \text{sec m$^{-1/3}$}) \sqrt[3]{m}
\; \simeq \; (1.13 \times 10^{5}\, \text{y m$^{-1/3}$}) \sqrt[3]{m}.
\end{equation}
The time $t_{2}$ defines the time interval $[t_1,t_2]$ during which
the classical regime holds. As we see, for macroscopic systems this
is a very long time, much longer than the time during which a
macro-object can be kept isolated from the rest of the universe, so
that its dynamics is described by Eq.~\eqref{eq:non-lin}.

To summarize, during the classical regime, which for macroscopic
systems lasts very long, the wave function behaves, for all
practical purposes, like a point moving deterministically in space
according to Newton's laws. In other words, the wave function
reproduces the motion of a classical particle. \vskip 0.3cm

\noindent \textsc{3. The diffusive regime.} After time $t_{2}$, two
new effects become dominant: First, the wave function converges
towards a Gaussian state, as we shall prove. Second, the motion
becomes more and more erratic: the dynamics begins to depart from
the classical one, showing its intrinsic stochastic nature. \vskip
0.3cm

A thorough mathematical analysis of these time regimes and their
main properties is still lacking. In this paper, as we have
anticipated, we focus now only on the long time behavior of the
solutions of Eq.~\eqref{eq:non-lin}, leaving the study of the
remaining properties as open problems for future research.

\section{Solution of the equation}
\label{sec:two}

In the first part of this section we derive the Green's
function~\eqref{eq:dfvc} in a way which will make clear the
connection between Eq.~\eqref{eq:lin} and the equation of the so
called non-self-adjoint (NSA) harmonic
oscillator~\cite{dav1,dav2,dav3}. This connection is important for
two reasons; from a physical point of view, it will bring a deep
insight on how the collapse of the wave function actually %
works. From a mathematical point of view, it will allow to prove
rigorously both the theorem 1 and 2 presented in the introductory
section.

A way to connect Eq.~\eqref{eq:lin} with that of the NSA harmonic
oscillator is to apply suitable transformations to the wave function
in such a way to transform the SDE in a Schr\"odinger-like equation.
We will do this in two steps. We present this section in detail for
convenience although the approach goes back to Kolokoltsov
\cite{kol4}. \vskip 0.3cm

\noindent \textsc{1. Reduction of Eq.~\eqref{eq:lin} to a linear
differential equation with random coefficients.} The idea is to
remove the stochastic differential term $\sqrt{\lambda} q d\xi_{t}$
from Eq.~\eqref{eq:lin}: borrowing the language of quantum
mechanics, we shift to a sort of interaction picture by defining a
suitable operator which maps the solution of Eq.~\eqref{eq:lin} to
the solution of a new equation which does not have that stochastic
term. To this end, let us consider the operator ${\mathcal Q}_{a}:
{\mathcal D}({\mathcal Q}_{a}) \subseteq {\mathcal L}^2({\mathbb R})
\rightarrow {\mathcal L}^2({\mathbb R})$ defined as follows:
\begin{equation} \label{eq:pos}
{\mathcal Q}_{a} \phi(x) \; = \; e^{ax}\phi(x), \qquad a \in
{\mathbb C};
\end{equation}
where ${\mathcal D}({\mathcal Q}_{a})$ is defined as the set of all
$\phi(x) \in {\mathcal L}^2({\mathbb R})$ such that $e^{ax}\phi(x)
\in {\mathcal L}^2({\mathbb R})$. It should be noted that, in
general, the operator ${\mathcal Q}_{a}$ is unbounded and its domain
${\mathcal D}({\mathcal Q}_{a})$ is dense in ${\mathcal
L}^2({\mathbb R})$ but not coincide with it. We will settle al
technical issues in the second part of the section. We now define
the vector:
\begin{equation}
\phi^{\text{\tiny(1)}}_{t} \; = \; {\mathcal Q}_{-\sqrt{\lambda}
\xi_{t}}\, \phi_{t};
\end{equation}
an easy application of It\^o calculus shows that
$\phi^{\text{\tiny(1)}}_{t}$ satisfies the differential equation:
\begin{equation} \label{eq:for-phibar}
d \phi^{\text{\tiny(1)}}_{t} \; = \;  \left[ - \frac{i}{\hbar} \,
{\mathcal Q}_{-\sqrt{\lambda} \xi_{t}}\, \frac{p^2}{2m}\, {\mathcal
Q}_{-\sqrt{\lambda} \xi_{t}}^{-1} - \, \lambda\, q^2 \right]
\phi^{\text{\tiny(1)}}_{t} \, dt, \qquad\qquad
\phi^{\text{\tiny(1)}}_{0} = \phi.
\end{equation}
The stochastic differential $\sqrt{\lambda} q d\xi_{t}$ has
disappeared; in turn, the free Hamiltonian $p^2/2m$ has been
replaced by the operator ${\mathcal Q}_{-\sqrt{\lambda} \xi_{t}}\,
(p^2/2m)\, {\mathcal Q}_{-\sqrt{\lambda} \xi_{t}}^{-1}$ which, due
to the specific commutation relations between $q$ and $p$, takes the
simple form:
\begin{equation}
{\mathcal Q}_{-\sqrt{\lambda} \xi_{t}}\, p^2\, {\mathcal
Q}_{-\sqrt{\lambda} \xi_{t}}^{-1} \; = \; p^2 \; - \; 2i \hbar\,
\sqrt{\lambda}\, \xi_{t}\, p \; - \; \lambda\, \hbar^2 \xi^2_{t};
\end{equation}
Eq.~\eqref{eq:for-phibar} can then be re-written as follows:
\begin{equation} \label{eq:q-eq}
i\hbar\, \frac{d}{dt}\, \phi^{\text{\tiny(1)}}_{t} \; = \; \left[
\frac{p^2}{2m} - i\hbar \lambda\, q^2 - \frac{i\hbar}{m}\,
\sqrt{\lambda}\, \xi_{t}\, p \; - \; \frac{\lambda\, \hbar^2}{2m}\,
\xi^2_{t} \right] \phi^{\text{\tiny(1)}}_{t}.
\end{equation}
This is a standard differential equation with random coefficients;
note that the operator on the right hand side is not self-adjoint,
due to the presence of the second and third term. The last term of
Eq.~\eqref{eq:q-eq} is a multiple of the identity operator and can
be removed by defining:
\begin{equation}
\phi^{\text{\tiny(2)}}_{t} \; = \; \exp\left[ - \frac{i
\hbar\lambda}{2m} \int_{0}^{t} \xi^{2}_{s}\, ds \right]
\phi^{\text{\tiny(1)}}_{t};
\end{equation}
we then obtain:
\begin{equation} \label{eq:q-eq-bis}
i\hbar\, \frac{d}{dt}\, \phi^{\text{\tiny(2)}}_{t} \; = \; \left[
\frac{p^2}{2m} - i\hbar \lambda\, q^2 - \frac{i\hbar}{m}\,
\sqrt{\lambda}\, \xi_{t}\, p \right] \phi^{\text{\tiny(2)}}_{t}.
\end{equation}
The third term on the right-hand-side contains a time dependent
coefficient, and the next step aims at removing it.\vskip 0.3cm

\noindent \textsc{2. Reduction of Eq.~\eqref{eq:q-eq-bis} to a
differential equation with constant coefficients.} The idea we now
follow is to perform a transformation similar to a boost. We
introduce the operator ${\mathcal P}_{a}: {\mathcal D}({\mathcal
P}_{a}) \subseteq {\mathcal L}^2({\mathbb R}) \rightarrow {\mathcal
L}^2({\mathbb R})$ defined as:
\begin{equation} \label{eq:mom}
{\mathcal P}_{ia/\hbar}\phi(x) \; = \;  \phi(x + a), \qquad a \in
{\mathbb C},
\end{equation}
where ${\mathcal D}({\mathcal P}_{a})$ is the set of all $\phi(x)
\in {\mathcal L}^2({\mathbb R})$ which can be analytically continued
to the line $x+a$ in the complex space ${\mathbb C}$, and such that
$\phi(x+a) \in {\mathcal L}^2({\mathbb R})$. Similarly to ${\mathcal
Q}_{a}$, also ${\mathcal P}_{a}$ is in general an unbounded operator
and its domain ${\mathcal D}({\mathcal P}_{a})$, though being dense,
does not coincide with ${\mathcal L}^2({\mathbb R})$; we will come
back to this point later in this section. We define the operator:
\begin{equation}
{\mathcal V}_{t} \; = \; \exp\left(-\,i a_{t}/ \hbar \right)\,
{\mathcal P}_{i b_{t}/ \hbar} \, {\mathcal Q}_{-i c_{t}/ \hbar},
\end{equation}
where the coefficients $a_{t}$, $b_{t}$ and $c_{t}$, yet to be
determined, will turn out to be complex random functions of time.
One can easily verify that:
\begin{eqnarray}
{\mathcal V}_{t}\, q \, {\mathcal V}^{-1}_{t} & = & q \, + \, b_{t}, \\
{\mathcal V}_{t}\, p \, {\mathcal V}^{-1}_{t} & = & p \, + \, c_{t},
\end{eqnarray}
and similarly for higher powers of $q$ and $p$. Let us define the
vector:
\begin{equation}
\varphi_{t} \; = \; {\mathcal V}_{t} \, \phi^{\text{\tiny(2)}}_{t},
\end{equation}
which solves the equation:
\begin{eqnarray} \label{eq:boost}
i\hbar\, \frac{d}{dt}\, \varphi_{t} & = & \left[ \frac{p^2}{2m} \; -
\; i\hbar \lambda\, q^2 \; - \; \left(\dot{b}_{t} - \frac{1}{m}\,
c_{t} +
\frac{i\hbar}{m} \sqrt{\lambda}\, \xi_{t} \right) p + \right. \nonumber \\
& + & \left.\left( \dot{c}_{t} - 2 i\, \hbar\, \lambda\, b_{t}
\right) q \; + \; \left(\dot{a}_{t} + \dot{c}_{t}\, b_{t} +
\frac{1}{2m}\, c^{2}_{t} - \frac{i\hbar}{m} \sqrt{\lambda}\,
\xi_{t}\, c_{t} - i \hbar\, \lambda\, b^{2}_{t} \right) \right]
\varphi_{t}.
\end{eqnarray}
The time-dependent part of the equation can be removed by requiring
that $a_{t}$, $b_{t}$ and $c_{t}$ satisfy the first-order
differential equations:
\begin{equation} \label{eq:lin-sys}
\left\{
\begin{array}{ll} \displaystyle
m\dot{b}_{t} -  c_{t} \; = \; - i\hbar
\sqrt{\lambda}\, \xi_{t} & \quad\qquad b_{0} = 0, \\
\dot{c}_{t} - 2 i\, \hbar\, \lambda\, b_{t} \; = \; 0 & \quad\qquad
c_{0} = 0
\end{array}
\right.
\end{equation}
and
\begin{equation}
\dot{a}_{t} \; + \; i\hbar\, \lambda\, b^{2}_{t} \; + \;
\frac{1}{2m}\, c^{2}_{t} \; - \; \frac{i\hbar}{m} \sqrt{\lambda}\,
\xi_{t}\, c_{t} \quad = \quad 0, \quad\qquad a_{0} = 0.
\end{equation}
The first two equations form a non-homogeneous linear system of
first order differential equations, which has a unique ${\mathbb
Q}$-a.s. continuous random solution; the third equation instead
determines the global factor $a_{t}$, which is also random. With
such a choice for the three parameters, Eq.~\eqref{eq:boost}
becomes:
\begin{equation} \label{eq:final}
i\hbar\, \frac{d}{dt}\, \varphi_{t} \;  = \; \left[ \frac{p^2}{2m}
\; - \; i\hbar \lambda\, q^2 \right] \varphi_{t}, \qquad\qquad
\varphi_{0} = \phi,
\end{equation}
which is the equation of the so-called non-self-adjoint (NSA)
harmonic oscillator, whose solution and most important properties
are well known. Before continuing, we note that in the case of a
more general Hamiltonian $H = p^2/2m + V(q)$ appearing in
Eq.~\eqref{eq:non-lin} in place of just the free evolution $p^2/2m$,
the potential $V(q)$ would have been transformed, when going from
Eq.~\eqref{eq:q-eq-bis} to Eq.~\eqref{eq:final}, according to the
rule: ${\mathcal V}_{t} V(q) {\mathcal V}^{-1}_{t} = V(q + b_{t})$;
in this case, we would not be able to remove completely the
time-dependent terms from the equation and we would not be able to
reduce the original equation to one, whose solution is known.
However, besides the free particle case, all equations containing
terms at most quadratic in $q$ and $p$ (among them, the important
case of the harmonic oscillator) can be solved in a similar way.

The solution of Eq.~\eqref{eq:final} admit a representation in terms
of the Green's function:
\begin{equation}
G_{t}^{\text{\tiny NSA}}(x,y) \; = \; \sqrt{\frac{\lambda}{\upsilon
\pi \sinh \upsilon t}} \exp \left[ - \frac{\lambda}{\upsilon} (x^2 +
y^2)\coth \upsilon t + 2\, \frac{\lambda}{\upsilon}\, x\, y
\sinh^{-1} \upsilon t \right],
\end{equation}
with $\upsilon$ and $\omega$ defined as in~\eqref{eq:gfdsp}. In this
way we have established the link between the solutions of the
SDE~\eqref{eq:lin} and those of the equation for the NSA harmonic
oscillator~\eqref{eq:final}, which we summarize in the following
lemma, whose proof is straightforward. \vskip 0.3cm

\noindent \textsc{Lemma~\ref{sec:two}.1:} Let ${\mathcal
T}^{\text{\tiny NSA}}_{t}$ be the evolution operator represented by
the Green's function $G_{t}^{\text{\tiny NSA}}(x,y)$ and ${\mathcal
T}_{t}$ the one represented by $G_{t}(x,y)$; then:
\begin{equation} \label{eq:gen-sol}
{\mathcal T}_{t} \; \equiv \; \exp\left(i
\vartheta_{t}/\hbar\right)\, {\mathcal Q}_{\sqrt{\lambda} \xi_{t} +
(i c_{t} / \hbar)} \, {\mathcal P}_{-i b_{t} /\hbar} \, {\mathcal
T}^{\text{\tiny NSA}}_{t},
\end{equation}
where the two random functions $b_{t}$ and $c_{t}$ solve the linear
system~\eqref{eq:lin-sys}, and $\vartheta_{t}$, which includes all
global, i.e. independent of $x$, phase factors, solves the equation:
\begin{equation} \label{eq:theta}
\dot{\vartheta}_{t} \quad = \quad - \, i\hbar\, \lambda\, b^{2}_{t}
\, - \, \frac{1}{2m}\, c^{2}_{t} \, + \, \frac{i\hbar}{m}
\sqrt{\lambda}\, \xi_{t}\, c_{t} + \frac{\lambda \hbar^2}{2m}
\xi^{2}_{t}, \quad\qquad \theta_{0} = 0.
\end{equation} \vskip 0.3cm

We now proceed to prove in which sense $\phi_{t} := {\mathcal
T}_{t}\phi$ is the topological strong solution of Eq.~\eqref{eq:lin}
for the given initial condition $\phi$. We first need to set some
properties of the Green's function $G_{t}^{\text{\tiny NSA}}(x,y)$
which will be necessary for the subsequent theorem. \vskip 0.3cm

\noindent \textsc{Lemma~\ref{sec:two}.2:} The absolute value of
$G_{t}^{\text{\tiny NSA}}(x,y)$ is equal to:
\begin{equation}
| G_{t}^{\text{\tiny NSA}}(x,y) | \; = \; \sqrt{\frac{2 \lambda}{\pi
\omega \sqrt{\cosh \omega t - \cos \omega t}}} \exp \left[ -
\frac{\lambda}{\omega}\, (x^2 + y^2)\, p_{t} + 4\,
\frac{\lambda}{\omega}\, x \, y\, q_{t} \right],
\end{equation}
where we have introduced the following quantities:
\begin{eqnarray}
p_{t} & = & \frac{\sinh \omega t - \sin \omega t}{\cosh \omega t -
\cos \omega t}, \label{eq:dgds5} \\
q_{t} & = & \frac{\sinh \omega t/2 \cos \omega t/2 - \cosh \omega
t/2 \sin \omega t/2}{\cosh \omega t - \cos \omega t};
\end{eqnarray}
note that the function $p_{t}$ is positive for any $t > 0$.
The integral of $|G_{t}^{\text{\tiny NSA}}(x,y)|^2$ with respect to
$y$ is equal to:
\begin{equation} \label{eq:prop3}
\int dy \, | G_{t}^{\text{\tiny NSA}}(x,y) |^2 \; = \; \sqrt{\frac{2
\lambda}{\pi \omega (\sinh \omega t - \sin \omega t)}} \exp \left[ -
2 \, \frac{\lambda}{\omega}\, \frac{p^{2}_{t} - 4
q^{2}_{t}}{p_{t}}\, x^2 \right].
\end{equation}
A simple calculation shows that $p^{2}_{t} - 4 q^{2}_{t} > 0$ for
any $t > 0$; this means that $G_{t}^{\text{\tiny NSA}}(x,\cdot)$,
taken as a function of $y$, belongs to ${\mathcal L}^{2}({\mathbb
R})$ for any $x \in {\mathbb R}$ and $t > 0$; moreover:
\begin{equation} \label{eq:fineq}
\int dx \, \| G_{t}^{\text{\tiny NSA}}(x, \cdot) \|^2 \, < \, +
\infty \qquad\qquad \text{for any $t > 0$}.
\end{equation} \\
Finally, the following expression holds true:
\begin{eqnarray} \label{eq:prop4}
\int dy\, |e^{bx} G_{t}^{\text{\tiny NSA}}(x+a, y) |^2 & = &
\sqrt{\frac{2\lambda}{\pi \omega (\sinh \omega t - \sin \omega t)}}
\exp \left\{ - 2 \frac{\lambda}{\omega} \left[ \frac{p^{2}_{t} - 4
q^{2}_{t}}{p_{t}}\, x^2 \right. \right. \nonumber \\
& + & 2 \left( p_{t} a_{\makebox{\tiny R}} + \overline{p}_{t}
a_{\makebox{\tiny I}} - 4 \frac{q_{t}(q_{t}a_{\makebox{\tiny R}} +
\overline{q}_{t} a_{\makebox{\tiny I}})}{p_{t}}\, + \,
2b_{\makebox{\tiny R}} \right) x \nonumber
\\
& + & \left. \left. p_{t}(a_{\makebox{\tiny R}}^2 -
a_{\makebox{\tiny I}}^2) + 2 \overline{p}_{t} a_{\makebox{\tiny R}}
a_{\makebox{\tiny I}} - 4 \frac{(q_{t} a_{\makebox{\tiny R}} +
\overline{q}_{t} a_{\makebox{\tiny I}})^2}{p_{t}} \right] \right\},
\end{eqnarray}
with
\begin{eqnarray}
\overline{p}_{t} & = & \frac{\sinh \omega t + \sin \omega t}{\cosh
\omega t - \cos \omega t}, \\
\overline{q}_{t} & = & \frac{\sinh \omega t/2 \cos \omega t/2 +
\cosh \omega t/2 \sin \omega t/2}{\cosh \omega t - \cos \omega t}.
\end{eqnarray}
The above formulas imply that, for any $a, b \in {\mathbb C}$, for
any $x \in {\mathbb R}$ and for any $t > 0$, the function
$e^{bx}G_{t}^{\text{\tiny NSA}}(x+a, \cdot)$ belongs to ${\mathcal
L}^{2}({\mathbb R})$ and:
\begin{equation} \label{eq:ifin2}
\int dx \, \| e^{bx}G_{t}^{\text{\tiny NSA}}(x+a, \cdot) \|^2 \, <
\, + \infty.
\end{equation}

We are now in a position to state and prove the main theorem of this
section. \vskip 0.3cm

\noindent \textsc{Theorem~\ref{sec:two}.1:} Let ${\mathcal P}_{a}$
and ${\mathcal Q}_{a}$ be defined, respectively, as
in~\eqref{eq:mom} and~\eqref{eq:pos}; let $b_{t}$ and $c_{t}$ solve
the linear system~\eqref{eq:lin-sys} and $\theta_{t}$ be the
solution of Eq.~\eqref{eq:theta}. Finally, let $\phi_{t} = {\mathcal
T}_{t}\phi$, with $\phi \in {\mathcal L}^{2}({\mathbb R})$ and
${\mathcal T}_{t}$ defined as in~\eqref{eq:gen-sol}. Then the
following three statements hold true with probability 1:
\begin{eqnarray}
1. & & \mathcal{T}_t:\mathcal{L}^2(\mathbb{R}) \rightarrow \mathcal{L}^2(\mathbb{R}) \;\text{defines a bounded operator for every}\; t>0
\\
2. & & \phi \, \in \, {\mathcal L}^{2}_{B}({\mathbb R}) \quad
\Rightarrow \quad \phi_{t} \; \text{is a topological strong solution of Eq.~\eqref{eq:lin}}\\
3. & & \phi \, \in \, {\mathcal L}^{2}({\mathbb R}) \quad
\Rightarrow \quad \lim_{t \rightarrow 0} \| \phi_{t} - \phi \| \, =
0.
\end{eqnarray}
\vskip 0.3cm

\noindent \textsc{Proof of statement 1.} Let $\phi$ belong to
${\mathcal L}^{2}({\mathbb R})$; since also $G^{\text{\tiny
NSA}}_{t}(x, \cdot)$ belongs to ${\mathcal L}^{2}({\mathbb R})$ for
any $x \in {\mathbb R}$ and $t > 0$, H\"older's inequality implies
that $G_{t}^{\text{\tiny NSA}}(x,\cdot) \phi$ belongs to ${\mathcal
L}^{1}({\mathbb R})$; accordingly, the operator ${\mathcal
T}_{t}^{\text{\tiny NSA}}$ is well defined for any $t > 0$, and maps
any ${\mathcal L}^{2}({\mathbb R})$-function into a measurable
function. By using Schwartz inequality together with
relation~\eqref{eq:fineq}, we have:
\begin{equation} \label{eq:sch}
\int dx \left| \int dy \, G_{t}^{\text{\tiny NSA}}(x,y) \phi(y)
\right|^2 \; \leq \; \| \phi \|^{2} \int dx \, \| G_{t}^{\text{\tiny
NSA}}(x, \cdot) \|^{2} \; < \; + \infty;
\end{equation}
thus ${\mathcal T}_{t}^{\text{\tiny NSA}} \, \phi$ belongs to
${\mathcal L}^{2}({\mathbb R})$ for any $\phi$ in ${\mathcal
L}^{2}({\mathbb R})$ and for any $t > 0$.

In a similar way, since also $G_{t}^{\text{\tiny NSA}}(x+a, \cdot)$
belongs to ${\mathcal L}^{2}({\mathbb R})$ for any $a \in {\mathbb
C}$ and because of~\eqref{eq:ifin2}, one proves that ${\mathcal
P}_{a}{\mathcal T}_{t}^{\text{\tiny NSA}}\, \phi$ belongs to
${\mathcal L}^{2}({\mathbb R})$ for any $\phi \in {\mathcal
L}^{2}({\mathbb R})$, for any complex $a$ and for any $t > 0$, i.e.
that ${\mathcal D}({\mathcal P}_{a})$ contains ${\mathcal
R}({\mathcal T}_{t}^{\text{\tiny NSA}})$. Using once more the same
inequalities and~\eqref{eq:ifin2}, one shows also that ${\mathcal
Q}_{b}\,{\mathcal P}_{a}\, {\mathcal T_{t}^{\text{\tiny NSA}}}\,
\phi$ belongs to ${\mathcal L}^{2}({\mathbb R})$ for any $\phi$ in
${\mathcal L}^{2}({\mathbb R})$, fr any $a, b \in {\mathbb C}$ and
$t > 0$. \vskip 0.3cm

\noindent \textsc{Remark}: Actually a stronger statement is true, as
can be readily seen from the Gaussian form of the Green's function
$G_t$ of the operator $\mathcal{T}_t$: For positive $t$ it maps
${\mathcal L}^{2}({\mathbb R})$ to Schwartz space
$\mathcal{S}({\mathbb R})$. We shall need this information in the
proof of statement 3. \vskip 0.3cm

\noindent \textsc{Proof of statement 2.} Let us consider the vector
$\varphi_{t} := {\mathcal T_{t}^{\text{\tiny NSA}}} \phi$, with
$\phi \in {\mathcal L}^{2}_{B}({\mathbb R})$. By construction,
$\varphi_{t}$ solves Eq.~\eqref{eq:final}, once one proves that the
integration
\begin{equation} \label{eq:comm}
\int dy \, G_{t}^{\text{\tiny NSA}}(x, y) \phi(y)
\end{equation}
can be exchanged with the first and second partial derivatives with
respect to $x$ and with the first partial derivative with respect to
$t$. We note that the function $G_{t}^{\text{\tiny NSA}}(x, y)
\phi(y)$ satisfies the following two properties: i) The function $y
\mapsto G_{t}^{\text{\tiny NSA}}(x, y) \phi(y)$ is measurable and
integrable on ${\mathbb R}$ for any $t
>0$ and for any $x \in {\mathbb R}$; ii)
The first and second partial derivatives with respect to $x$ and the
first partial derivatives with respect to $t$ are exists for any
$t>0$, $x \in {\mathbb R}$ and $y \in {\mathbb R}$ and can be
bounded uniformly with respect to $t$ and $x$. Accordingly, one can
apply e.g. theorem 12.13 pag. 199 of~\cite{meas} to conclude that
the operations of integration and differentiation can be exchanged.


Having proved that $\varphi_{t}$ solve Eq.~\eqref{eq:final}, a
direct application of It\^o calculus proves that $\phi_{t}$, defined
as in~\eqref{eq:gen-sol}, is a topological strong solution of
Eq.~\eqref{eq:lin}. \vskip 0.3cm

\noindent \textsc{Proof of statement 3.} Let $\phi = \phi_0 \in
C^{\infty}_c(\mathbb{R})$ be given.
Since $\phi_{t}$ solves Eq.~\eqref{eq:lin} in a strong sense,
it also solves the SDE in a weak sense; hence, using e.g. Eq.~(1.1)
of~\cite{hol}, one has:
\begin{equation} \label{eq:cxzfdso}
\lim_{t \rightarrow 0} \, \langle \varphi | \phi_{t} \rangle \; = \;
\langle \varphi | \phi_{0} \rangle \qquad\quad \forall\,\, \varphi
\, \in \, C^{\infty}_c(\mathbb{R}).
\end{equation}
We extend~\eqref{eq:cxzfdso} to the general case of $\varphi  \in
{\mathcal L}^{2}({\mathbb R})$. Being dense in ${\mathcal
L}^{2}({\mathbb R})$, there exist a sequence $\{ \varphi_{n} \in
C^{\infty}_c(\mathbb{R}), n \in {\mathbb N } \}$ which approximates
any $\varphi \in {\mathcal L}^{2}({\mathbb R})$. By triangle and
Schwarz inequality we get
\begin{equation}\label{argument}
| \langle \varphi | \phi_{t} \rangle - \langle \varphi | \phi_{0}
\rangle | \; \leq \; | \langle \varphi_{n} | \phi_{t} \rangle -
\langle \varphi_{n} | \phi_{0} \rangle | + \| \varphi - \varphi_{n}
\| \| \phi_{t} \| + \| \varphi - \varphi_{n} \| \| \phi_{0} \|.
\end{equation}
The first term on the right-hand-side can be made arbitrarily small
because of~\eqref{eq:cxzfdso}; the second and third term can also be
made arbitrarily small by choosing $n$ sufficiently large, while $\|
\phi_{t} \|$ can be bounded as it converges to $\| \phi_{0} \|$ for
$t \rightarrow 0$, due to Eq.~\eqref{eq:hfgdsd4}. This proves that:
\begin{equation} \label{eq:cxzfdso2}
\lim_{t \rightarrow 0} \, \langle \varphi | \phi_{t} \rangle \; = \;
\langle \varphi | \phi_{0} \rangle \qquad\quad \forall\,\, \varphi
\, \in \, {\mathcal L}^{2}({\mathbb R}).
\end{equation}
Statement 3 for test functions $\phi \in C^{\infty}_c(\mathbb{R})$
now follows directly from Eq.~\eqref{eq:hfgdsd4},
Eq.~\eqref{eq:cxzfdso2} and observing $\| \phi_{t} - \phi_{0} \|^2 =
\| \phi_{t} \|^2 + \| \phi_{0} \|^2 -2 \langle \phi_{0} |\phi_{t}
\rangle^{\text{\tiny R}}$. It remains to extend the strong
continuity of $\mathcal{T}_t$ from the subspace
$C_c^{\infty}(\mathbb{R})$ to $\mathcal{L}^2(\mathbb{R})$.
For this observe that for $\phi \in C^{\infty}_c(\mathbb{R})$
$(\|\phi_t\|^2)_{t \geq 0}$ defines a stochastic process with
continuous paths and by Holevo's result (cf. Eq.~\eqref{eq:hfgdsd4})
it is a martingale. For given $f \in \mathcal{L}^2(\mathbb{R})$
choose a sequence $(\varphi^n)_{n \in \mathbb{N}} \subset
C^{\infty}_c(\mathbb{R})$, which converges to $f$ in
$\mathcal{L}^2(\mathbb{R})$. Doob's inequality for submartingales
implies that for all $n,m \in \mathbb{N}$, $T>0$ and $\lambda>0$
\begin{equation}\label{Doob}
\mathbb{Q}\left(\sup_{0 \leq t \leq
T}\bigl|\|\varphi^n_t\|^2-\|\varphi^m_t\|^2\bigr|>\lambda\right)
\leq
\frac{1}{\lambda}\mathbb{E}_{\mathbb{Q}}\bigl[|\|\varphi^n_T\|^2-\|\varphi^m_T\|^2\bigr|\bigr].
\end{equation}
We now show that
\begin{equation}\label{goal}
\lim_{n,m\rightarrow \infty}\mathbb{E}_{\mathbb
Q}\bigl[|\|\varphi^n_T\|^2-\|\varphi^m_T\|^2\bigr|\bigr] = 0.
\end{equation}
The elementary inequality
\begin{displaymath}
|\|\varphi^n_t\|_2^2-\|\varphi_t^m\|_2 ^2| \leq (\|\varphi_t^n\|_2 +
\|\varphi_t^m\|_2)\|\varphi_t^n-\varphi_t^m\|_2
\end{displaymath}
implies that
\begin{equation*}
\begin{split}
\mathbb{E}_{\mathbb Q}\bigl[|\|\varphi^n_t\|^2-\|\varphi_t^m\|^2| &
\leq \mathbb{E}_{\mathbb Q}\bigl[(\|\varphi_t^n\| +
\|\varphi_t^m\|)\|\varphi_t^n-
\varphi_t^m\|\bigr] \\
&\leq \bigl(\mathbb{E}_{\mathbb Q}\bigl[(\|\varphi_t^n\| +
\|\varphi_t^m\|)^2\bigr]\bigr)^{\frac{1}{2}}\bigl(\mathbb{E}_{\mathbb
Q}
\bigl[\|\varphi_t^n-\varphi_t^m\|^2\bigr]\bigr)^{\frac{1}{2}}\\
&\leq \sqrt{2}\bigl(\mathbb{E}_{\mathbb
Q}[\|\varphi_t^n\|^2]+\mathbb{E}_{\mathbb Q}
[\|\varphi_t^m\|^2]\bigr)^{\frac{1}{2}}\|\varphi^n-\varphi^m\|^2 \\
&=
\sqrt{2}\bigl(\|\varphi^n\|^2+\|\varphi^m\|^2\bigr)^{\frac{1}{2}}\|\varphi^n-\varphi^m\|.
\end{split}
\end{equation*}
The right hand side converges to $0$ as $n,m \rightarrow \infty$.
Therefore the sequence of stochastic processes
$(\|\varphi_t^n\|^2)_{t\geq 0}$ is a Cauchy sequence in the complete
metric space $(\mathbb{D},d)$ of adapted processes with right
continuous paths having left limits, where the metric $d$ is defined
as (see page 56 -- 57 in \cite{Pro} for background concerning this
topology)
\begin{displaymath}
d(X,Y)=\sum_{n=1}^{\infty}\frac{1}{2^n}\mathbb{E}_{\mathbb Q}
\left[\min\left(1,\sup_{0\leq s\leq
n}|(X-Y)_s|\right)\right]\,\,(X,Y \in \mathbb{D}).
\end{displaymath}
Therefore $(\|\varphi_t^n\|^2)_{t\geq 0}$ converges locally
uniformly in probability to a stochastic process. This stochastic
process again has to be continuous almost surely, since a
subsequence of $(\|\varphi_t^n\|^2)_{t\geq 0}$ converges locally
uniformly with probability one. Since $\lim_{n\rightarrow
\infty}\|\varphi_t^n\|^2 = \|f_t\|^2$ almost surely we know that
$[0,\infty)\ni t \mapsto \|f_t\|^2$ is continuous,
in particular $ \lim_{t\rightarrow 0} \|f_t\| = \|f\|$ almost surely and defines by the lemma of Fatou a positive continuous supermartingale.
Therefore it has a unique decomposition $\|f_t\|^2 = M_t - A_t$, where $(M_t)_{t \geq 0}$ is a continuous martingale and $(A_t)_{t \geq 0}$ is increasing process. In fact as we shall show now, the increasing process is identically $0$, i.e. $\|f_t\|_{t \geq 0}^2$ is a positive martingale for every $f \in \mathcal{L}^2(\mathbb{R})$.
For that we observed in the Remark above that for positive $\varepsilon$ the function $f_{\varepsilon}$ almost surely belongs to the Schwartz space and in particular to the domain of the generator. By Holevo's result cited above $(\|\mathcal{T}_{t-\varepsilon}f_{\varepsilon}\|)_{t \geq \varepsilon}$ is a continuous martingale. Therefore $A_t = 0$ for $t>0$ and hence it equals $0$ almost surely.
In order to ensure strong convergence $ \lim_{t \rightarrow \infty}\|f_t-f\| = 0$ we
need only show that weak convergence holds, i.e. $\lim_{t\rightarrow \infty}\langle\phi|f_t\rangle=\langle\phi|f\rangle$\,.
Observing
\begin{displaymath}
|\langle\psi|f_t\rangle-\langle\phi|f\rangle|\leq |\langle\phi|f_t\rangle-\langle\phi|\varphi^n_t\rangle|+|\langle\phi|\varphi_t^n\rangle-\langle\phi|\varphi^n\rangle|+|\langle\phi|\varphi^n\rangle
-\langle\phi|f\rangle|
\end{displaymath}
it suffices to show that for some $T>0$ $\lim_{n\rightarrow\infty}\sup_ {t \leq T}|\langle\phi|f_t\rangle-\langle\phi|\varphi^n_t\rangle|=0$.  But $ \sup_{t \leq T}|\langle\phi|f_t\rangle-\langle\phi|\varphi^n_t\rangle|\leq \|\phi\|\sup_{t\leq T}\|f_t-\varphi_t^n\|$. Therefore we need only establish that $\lim_{n  \rightarrow \infty}\sup_{t\leq T}\|f_t-\varphi_t^n\|=0$. This is done by a similar argument as above, namely we show that for every $\varepsilon > 0$
\begin{displaymath}
\lim_{n\rightarrow \infty}\mathbb{Q}\bigl(\sup_{t\leq T}\|f_t-\varphi_t^n\|^2>\varepsilon\bigr) = 0,
\end{displaymath}
because then there exists a subsequence which is almost surely convergent to $0$. But as we showed above $(\|g_t\|^2)_{t \geq 0}$ is a martingale for every $g  \in \mathcal{L}^2(\mathbb{R})$. Hence $(\|f_t-\varphi_t^n\|^2)_{t \geq 0}$ is a martingale and we can again apply Doob's inequality as before.
\vskip 0.3cm
\noindent \textsc{Remark 1.} The Gaussian form of the Green's
function~\eqref{eq:dfvc} is a consequence of the fact that
Eq.~\eqref{eq:lin} contains terms which are at most quadratic in $q$
and $p$. This also implies preserves shape of initially Gaussian
wave functions; in fact, as shown e.g. in~\cite{gg,kol3,kol4,ab1}, a
state
\begin{equation}
\phi_{t}(x) \; = \; \exp\left[ - \sigma_{t} (x - x^{\text{\tiny
m}}_{t})^2 + i k^{\text{\tiny m}}_{t}x + \varsigma_{t} \right],
\end{equation}
is solution of Eq.~\eqref{eq:lin} provided that the two real
parameters $x^{\text{\tiny m}}_{t}$, $k^{\text{\tiny m}}_{t}$ and
the two complex parameters $\sigma_{t}$, $\varsigma_{t}$ satisfy the
following stochastic differential equations:
\begin{eqnarray}
d \sigma_{t} & = & \left[ \lambda - \frac{2i\hbar}{m}\left(
\sigma_{t} \right)^2 \right]dt, \label{eq:gvf1}\\
d x^{\text{\tiny m}}_{t} & = & \frac{\hbar}{m} k^{\text{\tiny
m}}_{t} dt + \frac{\sqrt{\lambda}}{2 \sigma^{\text{\tiny R}}_{t}}
\left[ d \xi_{t} - 2 \sqrt{\lambda} x^{\text{\tiny m}}_{t}
dt \right], \label{eq:gvf2} \\
d k^{\text{\tiny m}}_{t} & = & - \sqrt{\lambda}
\frac{\sigma^{\text{\tiny I}}_{t}}{\sigma^{\text{\tiny R}}_{t}}
\left[ d \xi_{t} - 2 \sqrt{\lambda} x^{\text{\tiny m}}_{t} \right],
\label{eq:gvf3} \\
d \varsigma^{\text{\tiny R}}_{t} & = & \left( \lambda
(x^{\text{\tiny m}}_{t})^{2} + \frac{\hbar}{m}\,
\sigma^{\makebox{\tiny I}}_{t} + \frac{\lambda}{4
\sigma^{\makebox{\tiny R}}_{t}}\right)dt \; + \; \sqrt{\lambda}\,
x^{\text{\tiny m}}_{t} \left[ d\xi_{t} - 2\sqrt{\lambda}\,
x^{\text{\tiny m}}_{t}\, dt \right], \label{eq:gvf4} \\
d \varsigma^{\text{\tiny I}}_{t} & = & \left( - \frac{\hbar}{2m}\,
(k^{\text{\tiny m}}_{t})^{2} - \frac{\hbar}{m}\,
\sigma^{\makebox{\tiny R}}_{t} + \frac{\lambda
\sigma^{\makebox{\tiny I}}_{t}}{4 (\sigma^{\makebox{\tiny
R}}_{t})^2} \right)dt +  \sqrt{\lambda} \frac{\sigma^{\makebox{\tiny
I}}_{t}}{\sigma^{\makebox{\tiny R}}_{t}}\,x^{\text{\tiny m}}_{t}
\left[ d\xi_{t} - 2\sqrt{\lambda}\, x^{\text{\tiny m}}_{t}\, dt
\right].  \label{eq:gvf5}
\end{eqnarray}
In particular, the solution of Eq.~\eqref{eq:gvf1} is $\sigma_{t} =
(\lambda/\upsilon) \coth (\upsilon t + \kappa)$, where $\kappa$ sets
the initial condition. These results will be useful in the
subsequent analysis.

\section{Representation of the solution in terms of eigenstates
of the NSA harmonic oscillator}
\label{sec:three}

We now turn to the problem of analyzing the long time behavior of
the solution of the (norm-preserving) non-linear
Eq.~\eqref{eq:non-lin}. The representation of the solution
$\phi_{t}$ of Eq.~\eqref{eq:lin} in terms of the Green's
function~\eqref{eq:dfvc} is not suitable for controlling the long
time behavior; it turns out to be more convenient to express
$\phi_{t}$ in terms of the {\it eigenstates} of the NSA harmonic
oscillator, resorting to the connection which we previously
established between Eq.~\eqref{eq:lin} and~\eqref{eq:final}. In this
way, as we shall see, the collapse process will be manifest: the
coefficients of the superposition will decrease exponentially in
time, the damping being the faster, the higher the associated
eigenstate. Accordingly---when normalization is also taken into
account---in the large time limit only the ground state survives,
which has a Gaussian shape.

We first recall a few basic features of the Hamiltonian of the NSA
harmonic oscillator,
\begin{equation} \label{eq:nsa}
H \; \equiv \; \frac{p^2}{2m} \, - i \hbar \lambda q^2
\end{equation}
which has been studied in particular by Davies in a series of
papers~\cite{dav1,dav2} and reviewed in his recent book~\cite{dav3}.
The {\it eigenvalues} of $H$ are complex and equal to:
\begin{equation}
\lambda_{n} \; \equiv \; \frac{1-i}{2}\, \hbar \omega_{n}, \qquad
\omega_{n} \equiv \left(n + \frac{1}{2} \right) \, \omega,
\end{equation}
and the corresponding {\it eigenvectors} are:
\begin{equation} \label{eq:ev}
\phi^{(n)}(x) \; \equiv \; \sqrt{z}\, e^{-z^2 x^2/2}
\overline{H}_{n}(zx), \qquad z^2 \equiv (1-i)\, \sqrt{\frac{\lambda
m}{\hbar}}
\end{equation}
where $\overline{H}_{n}(x)$ is the normalized Hermite polynomial of
degree $n$. Since the argument of $\overline{H}_{n}$
in~\eqref{eq:ev} is complex, these eigenstates are not orthogonal;
it can be shown that they are linearly independent and form a
complete set, however they {\it do not} form a basis. As such, they
can not directly used to expand an initial state into a
superposition of the eigenstates of $H$. This problem can be
circumvented in the following way, also discussed by Davies.

It is easy to see that the sequences $\{ \phi^{(n)} \}$ and $\{
\phi^{(n)\star} \}$ form a bi-orthonormal system; one then defines
the (non-orthogonal) projection operators:
\begin{equation} \label{eq:oihghf}
P_{n}\phi \; \equiv \; \langle \phi | \phi^{(n)\star} \rangle \,
\phi^{(n)} \; = \; \alpha_{n} \phi^{(n)},
\end{equation}
which satisfy the relations:
\begin{equation} \label{eq:fhgdf}
P_{n}\, P_{m} \; = \; \delta_{n,m}\, P_{n}, \qquad \| P_{n} \| \; =
\; \| \phi^{(n)} \|^2 \qquad \text{and} \qquad \lim_{n \rightarrow
+\infty} \frac{\ln \| P_{n} \|}{n} = 2 c,
\end{equation}
where $c$ is an appropriate constant~\cite{dav3}. As we see,
although the states $\phi^{(n)}$ are normalized, in the sense that
\begin{equation}
\int_{-\infty}^{+\infty}  \phi^{(n)}(x) \phi^{(m)}(x)\, dx \; = \;
\delta_{n,m},
\end{equation}
the norm of the projection operators $P_{n}$ grows exponentially as
$n \rightarrow + \infty$. Finally, the following equality holds
true~\cite{dav3}:
\begin{equation} \label{eq:dav}
{\mathcal T}_{t}^{\text{\tiny NSA}} \; = \; \sum_{n=0}^{\infty}
e^{-(1+i)\omega_{n} t/2} P_{n} \qquad \text{for $t > 4c/\omega$}.
\end{equation}
A remarkable property of the above representation of the solution of
Eq.~\eqref{eq:final} in terms of the eigenstates of the
operator~\eqref{eq:nsa} is that it holds not for any $t \geq 0$, as
one would naively expect, but only for $t > 4c/\omega$. The reason
is that the norm of the projection operators $P_{n}$ grows
exponentially with $n$, so one has to wait for $t$ to be large
enough in order for the term $e^{-n \omega t /2}$ to suppress the
exponential growth of the projectors. From a physical point of view,
recalling the discussion of sec~\ref{sec:TimeEvo}, since the
constant $c$ is of order 1~\cite{dav3} and $\omega \simeq 5.01
\times 10^{-5}$ sec$^{-1}$, we see that the
representation~\eqref{eq:dav} holds true only in part of the
classical regime and in the diffusive regime, which is the one we
are interested in studying now, but not in the physically more
crucial collapse regime.

We now apply the above results to our problem; we will first proceed
in an informal way, and at the end we will prove the relevant
theorems. Let $\phi \in {\mathcal L}^{2}({\mathbb R})$; then,
according to~\eqref{eq:gen-sol} and~\eqref{eq:dav}:
\begin{eqnarray}
\phi_{t}(x) \; = \; {\mathcal T}_{t}\, \phi & = & e^{[\sqrt{\lambda}
\xi_{t} + ic_{t}/\hbar]x + i\vartheta_{t}/\hbar}
\sum_{n=0}^{+\infty} \alpha_{n} e^{-(1+i)\omega_{n} t/2}
\phi^{(n)}(x-b_{t})\;\;\;\;\;
\\
& = & e^{- z^2 (x - \overline{x}_{t})^2/2 + i \overline{k}_{t} x +
\gamma_{t}} \sqrt{z} \sum_{n=0}^{+\infty} \alpha_{n}
e^{-(1+i)\omega_{n} t/2}\, \overline{H}_{n}[z(x-b_{t})],
\label{eq:dsfg}
\end{eqnarray}
where $\alpha_{n} = \langle \phi | \phi^{(n)\star} \rangle$ (see
Eq.~\eqref{eq:oihghf}), while the two real parameters
$\overline{x}_{t}$, $\overline{k}_{t}$ and the complex parameter
$\gamma_{t}$ are defined as follows:
\begin{eqnarray}
\overline{x}_{t} & = & b^{\makebox{\tiny R}}_{t} + b^{\makebox{\tiny
I}}_{t} - (2/m \omega) c^{\makebox{\tiny
I}}_{t} + (\omega/2 \sqrt{\lambda}) \xi_{t}, \label{eq:gfhjgf}\\
\overline{k}_{t} & = & (m\omega/\hbar) b^{\makebox{\tiny I}}_{t} +
(1/\hbar)(c^{\makebox{\tiny R}}_{t} - c^{\makebox{\tiny I}}_{t}) +
\sqrt{\lambda} \xi_{t}, \\
\gamma_{t} & = & - (1 - i)(m\omega/4\hbar) (b_{t}^{2} -
\overline{x}_{t}^{2}) + (i/\hbar) \theta_{t}.
\end{eqnarray}
By resorting to Eqs.~\eqref{eq:lin-sys} and~\eqref{eq:theta}, and
after a rather long calculation, we obtain the following set of SDEs
for these parameters:
\begin{eqnarray}
d \overline{x}_{t} & = & \frac{\hbar}{m}\, \overline{k}_{t}\, dt \;
+ \; \sqrt{\frac{\hbar}{m}} \left[ d\xi_{t} \; - \; 2
\sqrt{\lambda} \overline{x}_{t} dt \right] \label{eq:prima}, \\
d \overline{k}_{t} & = & \sqrt{\lambda} \left[ d\xi_{t} \; - \; 2
\sqrt{\lambda} \overline{x}_{t} dt \right], \label{eq:seconda} \\
d \gamma^{\makebox{\tiny R}}_{t} & = &  \left[ \lambda
\overline{x}_{t}^{2} \; + \; \frac{\omega}{4} \right] dt \; + \;
\sqrt{\lambda} \overline{x}_{t} \left[ d\xi_{t} \; - \; 2
\sqrt{\lambda} \overline{x}_{t} dt \right], \label{eq:penultima}\\
d \gamma^{\makebox{\tiny I}}_{t} & = & - \left[ \frac{\hbar}{2m}
\overline{k}_{t}^{2} \; + \; \frac{\omega}{4} \right] dt \; - \;
\sqrt{\lambda} \overline{x}_{t} \left[ d\xi_{t} \; - \; 2
\sqrt{\lambda} \overline{x}_{t} dt \right]; \label{eq:ultima}
\end{eqnarray}
the initial conditions are: $\overline{x}_{0} = \overline{k}_{0} =
\gamma_{0} = 0$. Note that these equations are equivalent
to~\eqref{eq:gvf2}--\eqref{eq:gvf5}, with $\sigma_{t} =
\sigma_{\infty} = \lambda/\upsilon = z^2/2$, $\overline{x}_{t} =
x^{\text{\tiny m}}_{t}$, $\overline{k}_{t} = k^{\text{\tiny m}}_{t}$
and $\gamma_{t} = \varsigma_{t} + (1+i)\omega/4$; as a matter of
fact, the  above equations describe the time evolution (according to
Eq.~\eqref{eq:lin}) of the ground state of the NSA harmonic
oscillator, which is:
\begin{equation} \label{eq:dfdfsd7}
\phi^{\text{\tiny $\infty$}}_{t}(x) = \exp\left[- \frac{z^2}{2} (x -
\overline{x}_{t})^2 + i \overline{k}_{t} x + \gamma_{t}-
\frac{1+i}{4}\,\omega t \right], \qquad\quad \phi^{\text{\tiny
$\infty$}}_{0}(x) \; = \; \phi^{(0)}(x).
\end{equation}
As we shall prove in the next section, this is the state to
which---apart from normalization---any initial state converges to,
in the long time limit, hence the name $\phi^{\text{\tiny
$\infty$}}_{t}$.

As we see, due to the stochastic part of the dynamics, the argument
the Gaussian weighting factor and that of the Hermite polynomials of
Eq.~\eqref{eq:dsfg} are different functions of time, while for
analyzing the long time behavior of the wave function, it is more
convenient that both arguments display the same time dependence. We
thus modify the argument of the Hermite polynomials, to make it
equal to that of the weighting factor. To this end, let us define
$\zeta_{t} = \overline{x}_{t} - b_{t}$; we can then write:
\begin{eqnarray}
\overline{H}_{n}[z(x-b_{t})] & = & \frac{1}{\sqrt{\sqrt{\pi} 2^n
n!}}\,
H_{n}[z(x - \overline{x}_{t}) + z \zeta_{t}] \nonumber \\
& = & \frac{1}{\sqrt{\sqrt{\pi} 2^n n!}}\, \sum_{m=0}^{n}
\binom{n}{m}
(2 z \zeta_{t})^{n-m} H_{m}[z(x - \overline{x}_{t})] \nonumber \\
& = & \sum_{m=0}^{n} \frac{\sqrt{n!}}{\sqrt{m!}(n-m)!}\, (\sqrt{2}z
\zeta_{t})^{n-m} \overline{H}_{m}[z(x - \overline{x}_{t})],
\end{eqnarray}
where $H_{m}$ is the standard (not normalized) Hermite polynomial of
degree $m$; in going from the first to the second line, we have used
property~\eqref{eq:h3}. Resorting to the above relation, we can
rewrite Eq.~\eqref{eq:dsfg} as follows:
\begin{equation} \label{eq:fgbxcv}
\phi_{t}(x) \; = \; e^{i \overline{k}_{t} x + \gamma_{t} -
(1+i)\omega t/4} \sum_{m=0}^{+\infty}
\overline{\alpha}^{\makebox{\tiny $(m)$}}_{t} e^{-(1+i)m\omega t/2}
\phi^{(m)}(x - \overline{x}_{t});
\end{equation}
the functions $\phi^{(m)}$ are the eigenstates defined
in~\eqref{eq:ev}, while the time dependent coefficients
$\overline{\alpha}^{\makebox{\tiny $(m)$}}_{t}$ are defined as
follows:
\begin{equation} \label{eq:gfda}
\overline{\alpha}^{\makebox{\tiny $(m)$}}_{t} \; = \;
\sum_{k=0}^{+\infty} \alpha_{k+m}
\frac{\sqrt{(k+m)!}}{\sqrt{m!}k!}\, (\sqrt{2}z
\overline{\zeta}_{t})^k,
\end{equation}
where we have introduced the new quantity $\overline{\zeta}_{t}
\equiv e^{-(1+i)\omega t/2} \zeta_{t}$.

Eqs.~\eqref{eq:fgbxcv} and~\eqref{eq:gfda} represent the two main
formulas, which we will use in the next section to analyze the large
time behavior. Before doing this, we need to set these formulas on a
rigorous ground; we will do these with the following two lemmata.
\vskip 0.3cm

\noindent \textsc{Lemma~\ref{sec:three}.1:} Let $\phi \in {\mathcal
L}^{2}({\mathbb R})$ and $\alpha_{n} = \langle \phi |
\phi^{(n)\star} \rangle$, with $\phi^{(n)}$ defined as
in~\eqref{eq:ev}. Then the series~\eqref{eq:gfda} defining
$\overline{\alpha}^{\makebox{\tiny $(m)$}}_{t}$ is a.s. convergent
for any $m$ and any $t > 0$. Moreover, one has the following bound
on the coefficients:
\begin{equation} \label{eq:dfbcf}
\left| \overline{\alpha}^{\makebox{\tiny $(m)$}}_{t} \right| \; \leq
\; N_{t}\, e^{(c + 1/2)m}, \qquad \quad N_{t} \; \equiv \;
A\,\sum_{k=0}^{+\infty}\, \frac{e^{k(c +1) }|\sqrt{2}z
\overline{\zeta}_{t}|^k}{\sqrt{k^k}} \qquad\quad \text{a.s.},
\end{equation}
where $A$ is a constant independent of the Brownian motion
$\xi_{t}$. \vskip 0.3cm

\noindent \textsc{Proof:} Because of~\eqref{eq:fhgdf}, there exists
a constant $C_{1}$ such that:
\begin{equation} \label{eq:fdxv}
| \alpha_{n} | \; \leq \; \| \phi \| \| \phi^{(n)} \| \; = \; \|
\phi^{(n)} \| \; \leq \; C_{1} e^{n c}.
\end{equation}
Secondly, using Stirling formula, there exists a constant $C_{2}$
such that:
\begin{equation}
C_{2}^{-1} \sqrt{2\pi n} n^n e^{-n} \; < \; n! \; < \; C_{2}
\sqrt{2\pi n} n^n e^{-n},
\end{equation}
for $n>1$; we can then write the following estimate:
\begin{eqnarray}
\frac{\sqrt{(k+m)!}}{\sqrt{m!}k!} & \leq &
\frac{C_{2}^2}{\sqrt{2\pi}}\, \sqrt[4]{\frac{k+m}{mk^2}}\,
\frac{(k+m)^{(k+m)/2} e^{-(k+m)/2}}{m^{m/2} e^{-m/2} k^k e^{-k}}
\nonumber \\
& \leq & \frac{C_{2}^2}{\sqrt{\pi}}\, e^{-k(\ln k -2)/2 + m/2};
\label{eq:cxbcvb}
\end{eqnarray}
in the second line, we have used the inequality $(k+m) \ln (k+m)
\leq k \ln k + m \ln m + k + m$. Using Eqs.~\eqref{eq:fdxv}
and~\eqref{eq:cxbcvb}, we have the following bound:
\begin{equation}
\left| \alpha_{k+m} \frac{\sqrt{(k+m)!}}{\sqrt{m!}k!}\, (\sqrt{2} z
\overline{\zeta}_{t})^k \right| \; \leq \;
\frac{C_{1}C_{2}^{2}}{\sqrt[4]{\pi}}\, \frac{e^{k(c +1) }|\sqrt{2}z
\overline{\zeta}_{t}|^k}{\sqrt{k^k}}, \qquad k,m \geq 1.
\end{equation}
The cases $k = 0$ and $m = 0$ can be treated separately, giving the
same bound, with the only possible difference of an overall constant
factor. This proves convergence of the series defined
in~\eqref{eq:gfda} and the bound~\eqref{eq:dfbcf}. \vskip 0.3cm

\noindent \textsc{Theorem~\ref{sec:three}.1:} Let the conditions of
Lemma~\ref{sec:three}.1 be satisfied; let moreover
$\overline{\zeta}_{t} \equiv e^{-(1+i)\omega t/2} \zeta_{t}$, where
$\zeta_{t} = \overline{x}_{t} - b_{t}$ with $\overline{x}_{t}$ and
$b_{t}$ solutions of Eq.~\eqref{eq:prima} and~\eqref{eq:lin-sys},
respectively. Then the series defined in~\eqref{eq:fgbxcv} is a.s.
norm convergent for $t > \overline{t} \equiv (4c+1)/\omega$. In
addition, the following equality holds true:
\begin{equation} \label{eq:gxzz}
{\mathcal T}_{t}\, \phi \; = \; e^{i \overline{k}_{t} x + \gamma_{t}
- (1+i)\omega t/4} \sum_{m=0}^{+\infty}
\overline{\alpha}^{\makebox{\tiny $(m)$}}_{t} e^{-(1+i)m\omega t/2}
\phi^{(m)}(x - \overline{x}_{t}), \qquad t > \overline{t},
\end{equation}
where ${\mathcal T}_{t}$ is the evolution operator associated to the
Green's function~\eqref{eq:dfvc}. \vskip 0.3cm

\noindent \textsc{Proof:} According to~\eqref{eq:fhgdf}
and~\eqref{eq:dfbcf}, one has:
\begin{equation} \label{eq:sdfg6}
\left\| \overline{\alpha}^{\makebox{\tiny $(m)$}}_{t}
e^{-(1+i)m\omega t/2} \phi^{(m)}[z(x - \overline{x}_{t})] \right\|
\; \leq \; C_{1} N_{t} e^{(2c + 1/2 - \omega t/2)m},
\end{equation}
from which the conclusion follows. Comparing the two expressions of
Eq.~\eqref{eq:gen-sol} and  Eq.~\eqref{eq:fgbxcv} when the initial
state $\phi$ is an eigenstate $\phi^{(n)}$, we see that they
coincide on the dense subspace of all finite linear combinations of
$\phi^{(n)}$, and hence on the whole of ${\mathcal L}^{2}({\mathbb
R})$.

\section{The long time behavior}
\label{sec:four}

We are now in a position to study the long time behavior of the
solution of Eq.~\eqref{eq:non-lin}. Looking at
expressions~\eqref{eq:fgbxcv} for the solution $\phi_{t}$
and~\eqref{eq:gfda} for the coefficients
$\overline{\alpha}^{\makebox{\tiny $(m)$}}_{t}$, it should be clear
what the long time behavior of the normalized solution $\psi_{t} =
\phi_{t}/\| \phi_{t} \|$ is: whatever the initial condition, at any
time $t > 0$ the wave function $\phi_{t}$ picks up a component on
the ground state $\phi^{(0)}(x - \overline{x}_{t})$, since
$\overline{\alpha}^{\makebox{\tiny $(0)$}}_{t} \neq 0$ as long as at
least one of the coefficients $\alpha_{k}$ is not null, which is
always the case. Eq.~\eqref{eq:fgbxcv} on the other hand shows that
each term of the superposition has an exponential damping factor,
which is the bigger, the higher the eigenvalue. Accordingly, after
normalization, only the eigenstate with the weakest damping factor
survives, which is the ground state. Hence we expect that the
general solution of Eq.~\eqref{eq:non-lin} converges a.s., in the
large time limit, to the ground state $\phi^{(0)}(x -
\overline{x}_{t})$, which is a Gaussian state. That this is true is
proven in the following theorem.\vskip 0.3cm

\noindent \textsc{Theorem~\ref{sec:four}.1:} Let $\phi_{t}$ be a
strong solution of Eq.~\eqref{eq:lin} that admits, for $t >
\overline{t}$ a representation as in~\eqref{eq:gxzz}. Let $\psi_{t}
\equiv \phi_{t} / \| \phi_{t} \|$  (when $\| \phi_{t} \| \neq 0$),
which can be written as follows:
\begin{equation} \label{eq:fdsdsd5}
\psi_{t}  \; = \; \psi_{t}^{\text{\tiny $\infty$}} \; + \;
e^{i(\overline{k}_{t} x + \gamma^{\makebox{\tiny I}}_{t} - \omega t
/4)}\sum_{m=1}^{+\infty} \frac{\overline{\alpha}^{\makebox{\tiny
$(m)$}}_{t}}{r_{t}}\, e^{-(1+i)m\omega t/2} \phi_{m}(x -
\overline{x}_{t}),
\end{equation}
with:
\begin{eqnarray}
\psi_{t}^{\text{\tiny $\infty$}} & := &
\frac{\overline{\alpha}^{\makebox{\tiny $(0)$}}_{t}}{r_{t}}\,
e^{i(\overline{k}_{t} x + \gamma^{\makebox{\tiny I}}_{t} - \omega t
/4)} \phi_{0}(x - \overline{x}_{t}), \label{eq:dgh5}\\
r_{t} & := & \left\| \sum_{m=0}^{+\infty}
\overline{\alpha}^{\makebox{\tiny $(m)$}}_{t} e^{-(1+i)m\omega t/2}
\phi_{m}(x - \overline{x}_{t}) \right \|.\label{eq:fgdfd0}
\end{eqnarray}
Then, with ${\mathbb P}$-probability 1:
\begin{equation}
\lim_{t \rightarrow \infty} \| \psi_{t} - \psi^{\text{\tiny
$\infty$}}_{t} \| \; = \; 0.
\end{equation}
Note that, apart from global factors, $\psi_{t}^{\text{\tiny
$\infty$}}$ is the ground state of the NSA harmonic oscillator,
randomly displaced both in position space as well as in momentum
space. \vskip 0.3cm

\noindent \textsc{Proof.} According to Eq.~\eqref{eq:fdsdsd5}, all
we need to prove is that, with ${\mathbb P}$-probability 1:
\begin{equation}
\lim_{t \rightarrow \infty} \left\| \sum_{m=1}^{+\infty}
\frac{\overline{\alpha}^{\makebox{\tiny $(m)$}}_{t}}{r_{t}}\,
e^{-(1+i)m\omega t/2} \phi_{m}(x - \overline{x}_{t}) \right\| \; =
\; 0.
\end{equation}
Resorting to~\eqref{eq:sdfg6}, one can write the following bound:
\begin{equation}
\left\| \sum_{m=1}^{+\infty} \frac{\overline{\alpha}^{\makebox{\tiny
$(m)$}}_{t}}{r_{t}}\, e^{-(1+i)m\omega t/2} \phi_{m}(x -
\overline{x}_{t}) \right\| \; \leq \; C_{1} \frac{N_{t}}{r_{t}} \,
\frac{e^{-\omega(t-\overline{t})}}{1 - e^{-\omega(t-\overline{t})}},
\end{equation}
thus all we need to set is the long time behavior of $r_{t}$ and
$N_{t}$. Lemmas~\ref{sec:four}.1 and~\ref{sec:four}.2 (see
Eqs.~\eqref{eq:zpewnb} and~\eqref{eq:gywiox}) state that, with
${\mathbb P}$-probability 1, $r_{t}$ converges asymptotically to a
finite and non-null random variable, while $N_{t}$ converges to a
finite random variable. From these properties, the conclusion of the
theorem follows immediately. \vskip 0.3cm

In the remaining of the section, we prove the required lemmas.
\vskip 0.3cm

\noindent \textsc{Lemma~\ref{sec:four}.1:} Let $r_{t}$ be defined as
in~\eqref{eq:fgdfd0}. Then, with ${\mathbb P}$--probability 1,
\begin{equation} \label{eq:zpewnb}
\lim_{t \rightarrow \infty} r_{t} \; = \; r_{\infty}\;\;
\text{finite and not null}.
\end{equation}
\vskip 0.3cm

\noindent \textsc{Proof.} According to Eq.~\eqref{eq:fgbxcv}
and~\eqref{eq:fgdfd0}, the following equality holds:
\begin{equation}
\| \phi_{t} \| \; = \; e^{\gamma^{\makebox{\tiny R}}_{t} - \omega
t/4}\, r_{t};
\end{equation}
resorting to the stochastic differentials~\eqref{eq:hfgdsd4}
and~\eqref{eq:penultima} for $\| \phi_{t} \|^2$ and
$\gamma^{\makebox{\tiny R}}_{t}$ respectively, one can write down
the following stochastic differential equation for $r_{t}^{2}$:
\begin{equation}
d r_{t}^{2} \; = \; \left[ 2 \sqrt{\lambda}\, (\langle q \rangle_{t}
-  \overline{x}_{t})\, d \xi_{t} \, + \, 4 \lambda \, (
\overline{x}_{t}^2 - \langle q \rangle_{t}\,\overline{x}_{t})\, dt
\right] r_{t}^{2}, \qquad\quad r_{0}^2 = 1.
\end{equation}
By using relation~\eqref{eq:step2}, the above equation can be
re-written in terms of the Wiener process $W_{t}$ as follows:
\begin{equation}
d r_{t}^{2} \; = \; \left[ 2 \sqrt{\lambda} \, (\langle q
\rangle_{t} - \overline{x}_{t})\, d W_{t} \, + \, 4 \lambda \,
(\langle q \rangle_{t} - \overline{x}_{t})^2 \, dt \right]
r_{t}^{2}, \qquad\quad r_{0}^2 = 1,
\end{equation}
whose solution is:
\begin{equation} \label{eq:gzbvf}
r_{t}^{2} \; = \; \exp\left[ 2\sqrt{\lambda} \int_{0}^{t} (\langle q
\rangle_{s} - \overline{x}_{s})\, d W_{s} + 2 \lambda \int_{0}^{t}
(\langle q \rangle_{s} - \overline{x}_{s})^2 ds \right].
\end{equation}

The crucial point is to establish the behavior of the difference
$\langle q \rangle_{t} - \overline{x}_{t}$ between the mean position
of the general solution $\psi_{t}$ and the mean position of the
``asymptotic'' state $\psi^{\text{\tiny $\infty$}}_{t}$. Since
$\psi_{t}$ converges to $\psi^{\text{\tiny $\infty$}}_{t}$, we
expect $\langle q \rangle_{t} - \overline{x}_{t}$ to vanishes
asymptotically. That this is actually true with ${\mathbb
P}$-probability 1 is proven in Lemma~\ref{sec:four}.3 (see
Eq.~\eqref{eq:defh}), where indeed it is shown that the convergence
is exponentially fast. This fact, together with~\eqref{eq:gzbvf},
concludes the proof of the lemma. \vskip 0.3cm

\noindent \textsc{Lemma~\ref{sec:four}.2:} Let $N_{t}$ be defined as
in~\eqref{eq:dfbcf}. Then, with ${\mathbb P}$--probability 1,
\begin{equation} \label{eq:gywiox}
\lim_{t \rightarrow \infty} N_{t} \; = \; N_{\infty}\;\;
\text{finite}.
\end{equation}
\vskip 0.3cm

\noindent \textsc{Proof.} Looking back at Eq.~\eqref{eq:dfbcf}, we
see that in order to prove this lemma it is sufficient to show that
$\overline{\zeta}_{t}$ tends to a finite limit as $t \rightarrow
\infty$, with ${\mathbb P}$-probability 1. According to our previous
definition, $\overline{\zeta}_{t}$ is equal to:
\begin{equation}
\overline{\zeta}_{t} \; = \; e^{-(1+i)\omega t/2}(\overline{x}_{t} -
b_{t});
\end{equation}
Eqs.~\eqref{eq:lin-sys} and~\eqref{eq:gfhjgf}, together with the
change of measure~\eqref{eq:step2}, lead to the following stochastic
differential equation for $\overline{\zeta}_{t}$ in terms of the
Wiener process $W_{t}$:
\begin{equation}
d \overline{\zeta}_{t} \; = \; \frac{\omega}{2
\sqrt{\lambda}}e^{-(1+i)\omega t/2} \left[ dW_{t} + 2 \sqrt{\lambda}
(\langle q \rangle_{t} - \overline{x}_{t}) dt \right], \qquad\quad
\overline{\zeta}_{0} = 0.
\end{equation}
Once again, the large time behavior of $\langle q \rangle_{t} -
\overline{x}_{t}$ (see Eq.~\eqref{eq:defh}) yields the conclusion of
the lemma. \vskip 0.3cm

\noindent \textsc{Lemma~\ref{sec:four}.3:} Let $\langle q
\rangle_{t} \equiv \langle \psi_{t} | q | \psi_{t} \rangle$ and
$\overline{x}_{t}$ defined in~\eqref{eq:gfhjgf}. Then, with
${\mathbb P}$-probability 1:
\begin{equation} \label{eq:defh}
h_{t} \; \equiv \; \langle q \rangle_{t} - \overline{x}_{t} \; = \;
O(e^{- \omega t/2}).
\end{equation}
\vskip 0.3cm

\noindent \textsc{Proof.} Let us consider the Gaussian solution of
Eq.~\eqref{eq:lin}:
\begin{eqnarray}
\phi^{\text{\tiny G}}_{t}(x) \; \equiv \; G_{t}(x,0) & = & K_{t}
\exp \left[ - \frac{\alpha_{t}}{2}x^2 \, + \,
\overline{a}_{t}x \, + \, \overline{c}_{t}\right] \\
& = & K_{t} \exp \left[ - \frac{\alpha_{t}}{2} (x -
\overline{x}^{\text{\tiny G}}_{t})^2 \, + \, i
\overline{k}^{\text{\tiny G}}_{t} x \, + \, \tilde{c}_{t}\right]
\label{eq:jhdfs4}
\end{eqnarray}
where $G_{t}(x, y)$ is the Green's function defined
in~\eqref{eq:dfvc} and
\begin{equation} \label{eq:hdsd0}
\overline{x}^{\text{\tiny G}}_{t} =
\frac{\overline{a}_{t}^{\text{\tiny R}}}{\alpha_{t}^{\text{\tiny
R}}}, \quad\qquad \overline{k}^{\text{\tiny G}}_{t} =
\overline{a}_{t}^{\text{\tiny I}} - \frac{\alpha_{t}^{\text{\tiny
I}}}{\alpha_{t}^{\text{\tiny R}}} \overline{a}_{t}^{\text{\tiny R}},
\quad\qquad \tilde{c}_{t} = \overline{c}_{t} + \frac{\alpha_{t}}{2}
(\overline{x}^{\text{\tiny G}}_{t})^2.
\end{equation}
Note that $\overline{x}^{\text{\tiny G}}_{t}$ is the mean position
of the Gaussian state $\phi^{\text{\tiny G}}_{t}$, while $\hbar
\overline{k}^{\text{\tiny G}}_{t}$ is its average momentum.
Obviously we can write:
\begin{equation} \label{eq:dfg}
h_{t} \;  = \; (\langle q \rangle_{t} - \overline{x}^{\text{\tiny
G}}_{t}) + (\overline{x}^{\text{\tiny G}}_{t} - \overline{x}_{t});
\end{equation}
lemma~\ref{sec:kol}.1 proves that $\langle q \rangle_{t} -
\overline{x}^{\text{\tiny G}}_{t}$ has the required asymptotic
behavior (see Eq.~\eqref{eq:gfxd7}), so all we need to show is that
also $\overline{x}^{\text{\tiny G}}_{t} - \overline{x}_{t}$ behaves
as required. Lemma~\ref{sec:kol}.1 was first proven in~\cite{kol4};
for completeness, we reproduce it in Appendix~\ref{sec:kol},
adapting it to our notation. The proof of the lemma is instructive
because it makes clear why it is convenient to analyze $\langle q
\rangle_{t} - \overline{x}^{\text{\tiny G}}_{t}$ separately from
$\overline{x}^{\text{\tiny G}}_{t} - \overline{x}_{t}$.

By letting the ground state of the NSA harmonic oscillator evolve
according to the Green's function $G_{t}(x,y)$, one can express
$\overline{x}_{t}$ in terms of the
functions~\eqref{eq:sad1}--\eqref{eq:sad6}; a straightforward
calculation leads to the following result:
\begin{equation}
\overline{x}_{t} - \overline{x}^{\text{\tiny G}}_{t} \; = \;
\frac{\omega}{2\lambda}\left[ (p_{t}^{-1} -
1)\overline{a}_{t}^{\text{\tiny R}} - \left( \frac{\beta_{t}
\overline{b}_{t}}{\alpha_{t} + \alpha_{\infty}} \right)^{\text{\tiny
\!\!\!\!\!\! R}} \right],
\end{equation}
where $\alpha_{\infty} \equiv \lim_{t \rightarrow \infty} \alpha_{t}
= 2 \lambda/ \upsilon$. By inspecting expressions~\eqref{eq:dgds5}
and~\eqref{eq:sad3}, we recognize that $p_{t}^{-1} - 1 =
O(e^{-\omega t})$ and $| \beta_{t} | = O(e^{-\omega t /2})$, thus in
order to prove the lemma all we have to do is to control the long
time behavior of $\overline{a}_{t}$, which in turn sets the
asymptotic behavior of $\overline{b}_{t}$ through~\eqref{eq:sad5}.
Inverting Eq.~\eqref{eq:hdsd0} we get:
\begin{equation} \label{eq:iokm8}
\overline{a}_{t} \; = \; \alpha_{t}\, \overline{x}^{\text{\tiny
G}}_{t} + i \overline{k}^{\text{\tiny G}}_{t},
\end{equation}
thus we can control $\overline{a}_{t}$ by controlling
$\overline{x}^{\text{\tiny G}}_{t}$ and $\overline{k}^{\text{\tiny
G}}_{t}$. These two quantities, being the average position and
(modulo $\hbar$) average momentum of the Gaussian
solution~\eqref{eq:jhdfs4}, satisfy the stochastic differential
equations~\eqref{eq:gvf2} and~\eqref{eq:gvf3}, with $\alpha_{t}/2$
in place of $\sigma_{t}$. By using the change of
measure~\eqref{eq:step2}, we can re-express these equations in terms
of the Wiener process $W_{t}$ as follows:
\begin{eqnarray}
d \overline{x}^{\text{\tiny G}}_{t} & = & \left[
\frac{\hbar}{m}\overline{k}^{\text{\tiny G}}_{t} +
\frac{2\lambda}{\alpha_{t}^{\text{\tiny R}}}\, f_{t} \right] dt
+ \frac{\sqrt{\lambda}}{\alpha_{t}^{\text{\tiny R}}}\,  dW_{t},\\
d \overline{k}^{\text{\tiny G}}_{t} & = & - 2 \sqrt{\lambda}
\frac{\alpha_{t}^{\text{\tiny I}}}{\alpha_{t}^{\text{\tiny R}}}
f_{t} dt - \sqrt{\lambda} \frac{\alpha_{t}^{\text{\tiny
I}}}{\alpha_{t}^{\text{\tiny R}}} dW_{t},
\end{eqnarray}
with $f_{t} \equiv \langle q \rangle_{t} - \overline{x}^{\text{\tiny
G}}_{t}$. By integrating the second equation, by using the strong
law of large numbers applied to $W_{t}$, Eq.~\eqref{eq:gfxd7} for
$f_{t}$ and the fact that $\alpha_{t}$ has an asymptotic finite
limit, one can show that, with ${\mathbb P}$-probability 1, the
process $\overline{k}^{\text{\tiny G}}_{t}$ grows slower than $t^2$,
for $t \rightarrow \infty$. By integrating now the first equation,
and by using the same properties as before, one can show that
$\overline{x}^{\text{\tiny G}}_{t}$ grows slower than $t^3$, for $t
\rightarrow \infty$ and again with ${\mathbb P}$-probability 1.
According to Eq.~\eqref{eq:iokm8} and~\eqref{eq:sad5}, we then have,
with ${\mathbb P}$-probability 1:
\begin{equation} \label{eq:ffdaa}
\overline{a}_{t} \; = \; o(t^3) \;\; \text{as $t \rightarrow
\infty$}, \qquad\quad \lim_{t \rightarrow \infty}\overline{b}_{t} \;
= \; \overline{b}_{\infty} \;\; \text{finite}.
\end{equation}
This proves that $\overline{x}_{t} - \overline{x}^{\text{\tiny
G}}_{t}$ has the required asymptotic behavior, hence the conclusion
of the lemma. \vskip 0.3cm

In this way we have proven that any initial state is ${\mathbb
P}$-a.s. norm convergent to the Gaussian state~\eqref{eq:dgh5},
which can be written as follows:
\begin{equation} \label{eq:sfrtn5}
\psi^{\text{\tiny $\infty$}}_{t} \; \equiv \;
\sqrt[4]{\frac{\pi}{z^{2}_{\makebox{\tiny R}}}}\exp\left[ -
\frac{z^2}{2}(x - \overline{x}_{t})^2 + i \overline{k}_{t}x + i
\left( \gamma^{\makebox{\tiny I}}_{t} - \frac{\omega}{4} t \right)
\right],
\end{equation}
which has a fixed finite spread both in position and in momentum,
given by~\cite{ab1}:
\begin{eqnarray}
\Delta_{q} & = & \langle\psi^{\text{\tiny $\infty$}}_{t} | (q -
\overline{x}_{t})^2 | \psi^{\text{\tiny $\infty$}}_{t} \rangle^{1/2}
\;\;\; = \; \sqrt{\frac{\hbar}{m\omega}}, \label{eq:sdt} \\
\Delta_{p} & = & \langle\psi^{\text{\tiny $\infty$}}_{t} | (p -
\hbar\overline{k}_{t})^2 | \psi^{\text{\tiny $\infty$}}_{t}
\rangle^{1/2} \; = \; \sqrt{\frac{\hbar m\omega}{2}}.
\end{eqnarray}
This corresponds almost to the minimum allowed by Heisenberg's
uncertainty relations, as $\Delta_{q} \Delta_{p} = \hbar /
\sqrt{2}$. Note also that, the more massive the particle, the
smaller the spread in position of the asymptotic Gaussian state:
this is a well known effect of the localizing property of
Eq.~\eqref{eq:non-lin}. Finally, Eqs.~\eqref{eq:prima}
and~\eqref{eq:seconda}, together with the change of
measure~\eqref{eq:step2}, tell how the average position
$\overline{x}_{t}$ and momentum $\hbar \overline{k}_{t}$ evolve in
time, as a function of the Wiener process $W_{t}$:
\begin{eqnarray}
d \overline{x}_{t} & = & \frac{\hbar}{m}\, \overline{k}_{t}\, dt \;
+ \; \omega h_{t} dt \; + \; \frac{\omega}{2\sqrt{\lambda}} dW_{t}
\label{eq:prima1}, \\
d \overline{k}_{t} & = & 2 \lambda h_{t} dt + \sqrt{\lambda} dW_{t},
\label{eq:seconda1}
\end{eqnarray}
which imply that there exist two random variables $X$ and $K$ such
that~\cite{kol4}:
\begin{eqnarray}
\overline{x}_{t} & = & X + \frac{\hbar}{m} K\, t + \sqrt{\lambda}
\frac{\hbar}{m} \int_{0}^{t} W_{s} ds + \sqrt{\frac{\hbar}{m}} W_{t}
+ O(e^{-\omega t/2}),
\label{eq:prima2} \\
\overline{k}_{t} & = & K + \sqrt{\lambda} W_{t} + O(e^{-\omega
t/2}). \label{eq:seconda2}
\end{eqnarray}
These parameters fully describe the time evolution of the Gaussian
state~\eqref{eq:sfrtn5}.

\section{Conclusions and outlook}
\label{sec:end}

In section~\ref{sec:TimeEvo} we have spotted three interesting time
regimes during which the wave function, depending on the values of
the parameters $\lambda$ and $m$, evolves in a different way. In the
central sections of this paper we have analyzed the long time
behavior, which pertains to the third regime, the diffusive one.
There are many other properties of the solutions of
Eq.~\eqref{eq:non-lin} which deserve to be analyzed, and in this
conclusive section we would like to point out a number of
interesting open problems.\vskip 0.3cm

\noindent \textsc{I: Collapse regime.} Let $\ell$ be the length
which discriminates between a localized and a non-localized wave
function, i.e. such that, defining with $\Delta^{\psi} q$ the spread
in position of a wave function $\psi$, we say that $\psi$ is {\it
localized in space} whenever $\Delta^{\psi} q \leq \ell$. In our
case, we must take $\ell > \sqrt{\hbar/m \omega}$, where
$\sqrt{\hbar/m \omega}$ is the asymptotic spread (see
Eq.~\eqref{eq:sdt}).\vskip 0.3cm

\noindent \textsc{Problem I.1: collapse time.} Let $\psi_{t}$ be the
solution of Eq.~\eqref{eq:non-lin}, for a given initial condition
$\psi \in {\mathcal L}^{2}({\mathbb R})$ such that $\Delta^{\psi} q
> \ell$. Let us define the collapse time $T_{\text{\tiny
COL}}^{\psi}$ as the first time at which the wave function is
localized in space:
\begin{equation}
T_{\text{\tiny COL}}^{\psi} \; := \; \min\{t : \Delta^{\psi_{t}} q
\leq \ell \}.
\end{equation}

\noindent \textsc{Question I.1.1:} How is $T_{\text{\tiny
COL}}^{\psi}$ distributed, as a random variable? In particular, is
it finite with ${\mathbb P}$-probability 1, as we expect it to
be~\cite{ab1}? What are its mean ${\mathbb E}_{\mathbb
P}[T_{\text{\tiny COL}}^{\psi}]$ and variance ${\mathbb V}_{\mathbb
P}[T_{\text{\tiny COL}}^{\psi}]$?

\noindent \textsc{Question I.1.2:} How does $T_{\text{\tiny
COL}}^{\psi}$ depend on the initial spread $\Delta^{\psi} q$, as
well as on the parameters $\lambda$ and $m$?

\noindent \textsc{Question I.1.3:} What is the probability that, for
$t > T_{\text{\tiny COL}}^{\psi}$, the wave function de-localizes in
space, i.e. acquires a spread greater than $\ell$, namely
$\Delta^{\psi} q \geq \ell + \epsilon$, where $\epsilon$ is an
arbitrary positive quantity? This kind of analysis is important
because it gives a measure of how stable the localization process
is. \vskip 0.3cm

\noindent \textsc{Problem I.2: collapse probability.} Let
$\overline{\psi} := \psi_{t}$, for $t = {\mathbb E}_{\mathbb
P}[T_{\text{\tiny COL}}^{\psi}]$. Let $\overline{x} := \langle
\overline{\psi} | q | \overline{\psi} \rangle$ be the position of
the wave function at the average time at which it is localized in
space.

\noindent \textsc{Question I.2.1:} How is $\overline{x}$ distributed
as a random variable?

\noindent \textsc{Question I.2.2:} Let $p(x)$ be the probability
density of $\overline{x}$; let $| \psi(x) |^2$ be the collapses
probability density given by the Born probability rule. When does it
happen that
\begin{equation}
d(x) \; := \; | p(x) - | \psi(x) |^2 | \; \leq \; \delta,
\end{equation}
where $\delta$ is an appropriately small number? How does this
depend on the values of the parameters $\lambda$ and $m$? \vskip
0.3cm

\noindent \textsc{II: Classical regime.} in the classical regime,
the wave function is expected to move, on the average, like a
classical free particle. \vskip 0.3cm

\noindent \textsc{Problem II.1: classical motion.} Let
$\overline{q}_{t}$ and $\overline{p}_{t}$ be the (quantum) average
position and momentum of $\psi_{t}$. Let $t > T_{\text{\tiny
COL}}^{\psi}$.

\noindent \textsc{Question II.1.1:} How are $\overline{q}_{t}$ and
$\overline{p}_{t}$ distributed, as random variables? In particular,
what are their mean ${\mathbb E}_{\mathbb P}[\overline{q}_{t}], \,
{\mathbb E}_{\mathbb P}[\overline{q}_{t}]$ and variances ${\mathbb
V}_{\mathbb P}[\overline{q}_{t}], \, {\mathbb V}_{\mathbb
P}[\overline{q}_{t}]$?

\noindent \textsc{Question II.1.2:} How do they depend on the values
of $\lambda$ and $m$?

\noindent \textsc{Question II.1.3:} Let $T_{\text{\tiny
DIF}}^{\psi}$ be te time at which the motion departs from the
classical one
\begin{equation}
T_{\text{\tiny DIF}}^{\psi} \; := \; \min\{t : {\mathbb V}_{\mathbb
P}[\overline{q}_{t}] \geq \Lambda_{q} \; \vee \; {\mathbb
V}_{\mathbb P}[\overline{p}_{t}] \geq \Lambda_{p} \},
\end{equation}
where $\Lambda_{q}$ and $\Lambda_{p}$ are suitable parameters
measuring the fluctuations of the position and momentum,
respectively, of the wave function. How does $T_{\text{\tiny
DIF}}^{\psi}$ depend on the parameters of the model? \vskip 0.3cm

\noindent \noindent \textsc{III: Diffusive regime.} This regime
begins after $T_{\text{\tiny DIF}}^{\psi}$, and it has been analyzed
in this paper: as we have seen, the wave keeps diffusing in the
Hilbert space, eventually taking a Gaussian shape, as described in
Sec.~\ref{sec:four}.

\section*{Acknowledgments}

The work was supported by the EU grant No. MEIF CT 2003-500543 and
by DFG (Germany).

\appendix
\section{Properties of Hermite polynomials}

We list here the main properties of Hermite polynomials, which are
used in the paper. The primary definition of the Hermite polynomials
is
\begin{equation} \label{eq:h1}
H_{n}(z) \; = \; n! \sum_{m=0}^{\lfloor n/2 \rfloor} \frac{(-1)^{m}
(2z)^{n-2m}}{m!(n-2m)!},
\end{equation}
where $z$ is any complex number. These polynomials satisfy the
following addition rule
\begin{equation} \label{eq:h3}
H_{n} (z_{1} + z_{2}) \; = \; \sum_{m=0}^{n} \binom{n}{m}
(2z_{2})^{n-m}\, H_{m}(z_{1}).
\end{equation}
When the argument is real ($z = x \in {\mathbb R}$), they form an
orthogonal set with respect to the weight $\exp[-x^2]$; the
normalized Hermite polynomials are:
\begin{equation} \label{eq:h5}
\overline{H}_{n}(x) \; = \; \frac{1}{N_{n}}\, H_{n}(x), \qquad N_{n}
= \sqrt{\sqrt{\pi} 2^n n!}.
\end{equation}

\section{Lemma} \label{sec:kol}

\noindent \textsc{Lemma~\ref{sec:kol}.1:} Let $\phi \in {\mathcal
L}^{2}({\mathbb R})$, $\| \phi \| = 1$ and let $\phi_{t} = {\mathcal
T}_{t} \phi$. Then, with ${\mathbb P}$-probability 1:
\begin{equation} \label{eq:gfxd7}
f_{t} \; \equiv \; \langle q \rangle_{t} - \overline{x}^{\text{\tiny
G}}_{t} \; = \; O(e^{- \omega t/2}),
\end{equation}
where $\langle q \rangle_{t} = \langle \psi_{t} | q | \psi_{t}
\rangle$, and $\overline{x}^{\text{\tiny G}}_{t}$ has been defined
in~\eqref{eq:hdsd0}. \vskip 0.3cm

\noindent \textsc{Proof.} Using the expression~\eqref{eq:dfvc} for
$G_{t}(x,y)$ together with Schwartz inequality, we can derive the
following bound on $\phi_{t}$:
\begin{equation} \label{eq:ytiytt}
| \phi_{t}(x) |^2 \leq | K_{t} |^2
\sqrt{\frac{\pi}{\alpha_{t}^{\text{\tiny R}}}}\, \exp \left[ - 2
\frac{\lambda}{\omega} \frac{p_{t}^2 - 4 q_{t}^2}{p_{t}} x^2 + 2
\left( \overline{a}_{t}^{\text{\tiny R}} + 8
\frac{\overline{b}_{t}^{\text{\tiny R}}q_{t}}{p_{t}}\right)x + 2
\overline{c}_{t}^{\text{\tiny R}} +\frac{\omega}{2\lambda}
\frac{(\overline{b}_{t}^{\text{\tiny R}})^2}{p_{t}}\right],
\end{equation}
which holds for any $t > 0$. The above inequality implies that it is
sufficient to consider $\phi \in {\mathcal L}^{2}({\mathbb R})$ such
that:
\begin{equation} \label{eq:dsfd9}
| \phi(x) | \; \leq \; C\, e^{-A x^2},
\end{equation}
where $C$ and $A$ are random variables. A direct calculation leads
to the following expression for the quantum average $\langle
\phi_{t} | q | \phi_{t} \rangle$:
\begin{eqnarray} \label{eq:fdas}
\langle \phi_{t} | q | \phi_{t} \rangle & = & | K_{t} |^2
\sqrt{\frac{\pi}{\alpha_{t}^{\text{\tiny R}}}} \exp\left[
2\overline{c}_{t}^{\text{\tiny R}} +
\frac{(\overline{a}_{t}^{\text{\tiny R}})^2}{\alpha_{t}^{\text{\tiny
R}}} \right] \int d y_{1} d y_{2}\, \phi(y_{1}) \phi(y_{2})^{\star}
\left[ \frac{\beta_{t}y_{1} + \beta_{t}^{\star} y_{2}}{2
\alpha_{t}^{\text{\tiny R}}} + \frac{\overline{a}_{t}^{\text{\tiny
R}}}{\alpha_{t}^{\text{\tiny R}}} \right] \nonumber \\
& & \cdot \exp\left[ -\frac{1}{2}\left( \alpha_{t} -
\frac{\beta_{t}^2}{2\alpha_{t}^{\text{\tiny R}}} \right)y_{1}^{2}
-\frac{1}{2}\left( \alpha_{t}^{\star} - \frac{\beta_{t}^{\star
2}}{2\alpha_{t}^{\text{\tiny
R}}} \right)y_{2}^{2} \right] \nonumber \\
& & \cdot \exp \left[ \left( \overline{b}_{t} +
\frac{\beta_{t}\overline{a}_{t}^{\text{\tiny
R}}}{\alpha_{t}^{\text{\tiny R}}} \right) y_{1} + \left(
\overline{b}_{t}^{\star} +
\frac{\beta_{t}^{\star}\overline{a}_{t}^{\text{\tiny
R}}}{\alpha_{t}^{\text{\tiny R}}} \right) y_{2} + \frac{|
\beta_{t}|^2}{2\alpha_{t}^{\text{\tiny R}}}\, y_{1} y_{2} \right].
\end{eqnarray}
As we shall soon see, all exponential terms in the above expression
can be controlled. The crucial factors are the two within brackets:
the first term decays exponentially in time, since $\beta_{t} =
O(e^{-\omega t /2})$, while $\alpha_{t}$ has a finite asymptotic
limit; the term $\overline{a}_{t}^{\text{\tiny
R}}/\alpha_{t}^{\text{\tiny R}}$, instead, does not decay in time
(see the discussion in connection with the proof of
lemma~\ref{sec:four}.3). Since $\| \phi_{t} \|^2$ is equal to the
expression~\eqref{eq:fdas} without the terms in square brackets, and
because of~\eqref{eq:hdsd0}, we have that
\begin{eqnarray} \label{eq:fdas9}
f_{t} \, \| \phi_{t} \|^2 & = & \langle \phi_{t} | q | \phi_{t}
\rangle - \frac{\overline{a}_{t}^{\text{\tiny
R}}}{\alpha_{t}^{\text{\tiny R}}} \| \phi_{t} \|^2 \; = \\
& = & | K_{t} |^2 \sqrt{\frac{\pi}{\alpha_{t}^{\text{\tiny R}}}}
\exp\left[ 2\overline{c}_{t}^{\text{\tiny R}} +
\frac{(\overline{a}_{t}^{\text{\tiny R}})^2}{\alpha_{t}^{\text{\tiny
R}}} \right] \int d y_{1} d y_{2}\, \phi(y_{1}) \phi(y_{2})^{\star}
\left[ \frac{\beta_{t}y_{1} + \beta_{t}^{\star} y_{2}}{2
\alpha_{t}^{\text{\tiny R}}} \right] \nonumber \\
& & \cdot \exp\left[ -\frac{1}{2}\left( \alpha_{t} -
\frac{\beta_{t}^2}{2\alpha_{t}^{\text{\tiny R}}} \right)y_{1}^{2}
-\frac{1}{2}\left( \alpha_{t}^{\star} - \frac{\beta_{t}^{\star
2}}{2\alpha_{t}^{\text{\tiny
R}}} \right)y_{2}^{2} \right] \nonumber \\
& & \cdot \exp \left[ \left( \overline{b}_{t} +
\frac{\beta_{t}\overline{a}_{t}^{\text{\tiny
R}}}{\alpha_{t}^{\text{\tiny R}}} \right) y_{1} + \left(
\overline{b}_{t}^{\star} +
\frac{\beta_{t}^{\star}\overline{a}_{t}^{\text{\tiny
R}}}{\alpha_{t}^{\text{\tiny R}}} \right) y_{2} + \frac{|
\beta_{t}|^2}{2\alpha_{t}^{\text{\tiny R}}}\, y_{1} y_{2} \right].
\end{eqnarray}
According to the discussion above, we expect the quantity $f_{t} \,
\| \phi_{t} \|^2$ to decay exponentially in time, as we shall now
prove; this is the reason why, in proving lemma~\ref{sec:four}.3, it
was convenient to split the difference $h_{t}$ as done in
Eq.~\eqref{eq:dfg}.

Using the inequality $y_{1} y_{2} \leq (y_{1}^2 + y_{2}^2)/2$ we can
write:
\begin{eqnarray}
|f_{t}| \| \phi_{t} \|^2 & = & \left|\langle \phi_{t} | q | \phi_{t}
\rangle - \frac{\overline{a}_{t}^{\text{\tiny
R}}}{\alpha_{t}^{\text{\tiny R}}} \| \phi_{t} \|^2 \right| \\
& \leq & \frac{|\beta_{t}|}{2 \alpha_{t}^{\text{\tiny R}}}|K_{t}|^2
\sqrt{\frac{\pi}{\alpha_{t}^{\text{\tiny R}}}} \exp\left[
2\overline{c}_{t}^{\text{\tiny R}} +
\frac{(\overline{a}_{t}^{\text{\tiny R}})^2}{\alpha_{t}^{\text{\tiny
R}}} \right] \int d y_{1} d y_{2}\, |\phi(y_{1})|
|\phi(y_{2})|(|y_{1}| + |y_{2}|) g(y_{1}) g(y_{2}),\;\;\;\;
\end{eqnarray}
with:
\begin{equation}
g(y) \; \equiv \; \exp\left[ - \frac{1}{2} \left(
\alpha_{t}^{\text{\tiny R}} - \frac{(\beta_{t}^{\text{\tiny
R}})^2}{\alpha_{t}^{\text{\tiny R}}} \right)y^2 + \left(
\overline{b}_{t}^{\text{\tiny R}} + \frac{\beta_{t}^{\text{\tiny
R}}\overline{a}_{t}^{\text{\tiny R}}}{\alpha_{t}^{\text{\tiny
R}}}\right)y \right].
\end{equation}
Next, by using the inequality $g(y_{1}) + g(y_{2}) \leq (g(y_{1})^2
+ g(y_{1})^2)/2$ and the symmetry between $y_{1}$ and $y_{2}$, we
have:
\begin{equation}
|f_{t}| \| \phi_{t} \|^2 \leq \frac{|\beta_{t}|}{2
\alpha_{t}^{\text{\tiny R}}}|K_{t}|^2
\sqrt{\frac{\pi}{\alpha_{t}^{\text{\tiny R}}}} \exp\left[
2\overline{c}_{t}^{\text{\tiny R}} +
\frac{(\overline{a}_{t}^{\text{\tiny R}})^2}{\alpha_{t}^{\text{\tiny
R}}} \right] \int d y_{1} d y_{2}\, |\phi(y_{1})|
|\phi(y_{2})|(|y_{1}| + |y_{2}|) g(y_{1})^2.
\end{equation}

Now, a direct computation shows that
\begin{equation}
\| G_{t}(\cdot,y) \|^2 \; \equiv \; \int dx | G_{t}(x,y) |^2 \; = \;
|K_{t}|^2 \sqrt{\frac{\pi}{\alpha_{t}^{\text{\tiny R}}}} \exp\left[
2\overline{c}_{t}^{\text{\tiny R}} +
\frac{(\overline{a}_{t}^{\text{\tiny R}})^2}{\alpha_{t}^{\text{\tiny
R}}} \right]  g(y)^2;
\end{equation}
the key point is that, since $G_{t}(x,y)$ solves Eq.~\eqref{eq:lin},
then $\| G_{t}(\cdot,y) \|^2$ is a positive martingale with respect
to the measure ${\mathbb Q}$, for any value of $y$; we call
$\text{Mar}_{\mathbb Q}(t,y)$ this martingale. We can then write:
\begin{eqnarray}
|f_{t}| \| \phi_{t} \|^2 & \leq & \frac{|\beta_{t}|}{2
\alpha_{t}^{\text{\tiny R}}}  \int d y_{1} d y_{2}\, |\phi(y_{1})|
|\phi(y_{2})|(|y_{1}| + |y_{2}|) \text{Mar}_{\mathbb Q}(t,y)
\nonumber \\
& \leq & \frac{|\beta_{t}|}{2 \alpha_{t}^{\text{\tiny R}}} \int dy
\, e^{-Ay^2} (A_{1} |y| + A_{2}) \text{Mar}_{\mathbb Q}(t,y),
\end{eqnarray}
where $A_{1}$ and $A_{2}$ are suitable constants. In going from the
first to the second line, we have used~\eqref{eq:dsfd9}. The
quantity
\begin{equation}
\frac{1}{2 \alpha_{t}^{\text{\tiny R}}} \int dy \, e^{-Ay^2} (A_{1}
|y| + A_{2}) \text{Mar}_{\mathbb Q}(t,y)
\end{equation}
is another positive martingale with respect to ${\mathbb Q}$, which
we call $\text{Mar}_{\mathbb Q}'(t)$. We arrive in this way at the
inequality:
\begin{equation}
|f_{t}| \; \leq \; |\beta_{t}| \, \frac{\text{Mar}_{\mathbb
Q}'(t)}{\| \phi_{t} \|^2}.
\end{equation}
Since $\text{Mar}_{\mathbb Q}'(t)$ is a positive martingale with
respect to ${\mathbb Q}$, then $\text{Mar}_{\mathbb P}(t) =
\text{Mar}_{\mathbb Q}'(t)/\| \phi_{t} \|^2$ is a positive
martingale with respect to ${\mathbb P}$ which, by Doob's
convergence theorem, has a ${\mathbb P}$-a.s. finite limit for $t
\rightarrow + \infty$. The conclusion of the lemma then follows from
Eq.~\eqref{eq:sad3}, according to which $\beta_{t} = O(e^{-\omega
t/2})$.

\end{document}